\documentclass[
preprint,
5p,
twocolumn,
]{elsarticle}
\usepackage{graphicx}
\usepackage{dcolumn}
\usepackage{bm}
\usepackage{amsmath,amssymb}
\usepackage{mathtools, cuted}

\usepackage{url}
\renewcommand{\url}[1]{}
\newcommand{\doi}[1]{}

\allowdisplaybreaks
\usepackage{subfigure}
\usepackage{subcaption}

\usepackage{float}
\usepackage[section]{placeins} 
\usepackage{parskip}
\usepackage[font=small,labelfont=bf]{caption}
\usepackage{setspace}
\usepackage{wrapfig}

\begin{document}
\title{Computational Methods of Wave Propagation for Semiclassical Models of High Harmonic Generation in Bulk Solids}
\author[1]{Ava N. Hejazi}
\author[2]{Nicholas Karpowicz}
\author[1]{Gregory D. Scholes}
\author[3]{Julia M. Mikhailova\corref{cor1}}
\affiliation[1]{Department of Chemistry at Princeton University}
\affiliation[2]{Max Planck Institute For Quantum Optics}
\affiliation[3]{Department of Mechanical and Aerospace Engineering at Princeton University}
\cortext[cor1]{Corresponding author}
\ead{j.mikhailova@princeton.edu}
\begin{abstract} We present a theoretical framework for self consistent treatment of nonlinear light-matter interactions in the ultrafast strong-field regime based on numerical solution of Maxwell's equations and semiconductor Bloch equations. This framework is shown to describe high-order harmonic generation and propagation in bulk semiconductors, investigating differences in reflected and transmitted harmonic spectra due to propagation effects. We show that the propagation of the combined field of the driving laser pulse and generated harmonics in a bulk semiconductor significantly modifies the harmonic spectra, affecting interpretation of experimental results relating the transmitted harmonic spectra to the underlying electronic structure of the material. This model allows the self-consistent description of strong-field light-matter interactions in the non-perturbative regime, opening the way to explore the transition between fully classical and quantum regimes of interaction, tunneling and multiphoton regimes of material ionization, and perturbative and non-perturbative regimes of harmonic generation in bulk materials. 
\end{abstract}

\maketitle
\thispagestyle{plain}
\section{Introduction}
High harmonic generation (HHG) is a strong field nonlinear optical process with applications in ultrafast spectroscopy, materials science, and metrology \cite{Ghimire2019High-harmonicSolids}. HHG in solids was first experimentally demonstrated in 2011 in 500 $\mu m$-thick ZnO crystals \cite{Ghimire2011ObservationCrystal}, and has since been widely studied in a range of semiconductor and insulating materials, as well as 2D and topological materials with novel electronic structure \cite{Kim2022TheoryPicture, Bai2021High-harmonicStates, Uzan2020AttosecondGeneration}. HHG spectra generated from solids are sensitive to the microscopic electronic structure, namely the variation of the band dispersions and transition dipole moments over the Brillouin zone \cite{Tancogne-Dejean2017ImpactSolids}. HHG thus offers a possible method of reconstructing the electronic structure of novel materials \cite{Vampa2015All-OpticalStructure, Wang2016DeterminationCrystal, Stepanov2017MappingSolids}. Similarly, this relationship between the electronic structure and the HHG spectra suggests that appropriate material engineering could help optimize attosecond XUV pulse generation for spectroscopic applications. 

The complex nature of the many competing nonlinear processes that may occur in materials under such high fields, however, means that propagation effects may obscure details of the microscopic mechanism and thereby necessitate accurate modeling and simulation methods to interpret experimental data and provide physical intuition. Numerical pulse propagation simulations in the regime of perturbative nonlinear optics \cite{Couairon2011PractitionersSimulation, Brabec1997NonlinearRegime, Kolesik2004NonlinearEquations, Karpowicz2023OpenSimulation} aim to capture the interplay of wave evolution with the microscopic polarization as described by phenomenological nonlinear susceptibility tensors \cite{Boyd2020NonlinearOptics}. The crucial role of phase-matching on perturbative harmonic generation efficiency in solids is one example of the intrinsic importance of wave propagation in conventional nonlinear optics. Phase-matching has also been shown to play a pivotal role in HHG efficiency in gases \cite{Niu2025ComprehensiveGases,LHuillier1991Higher-orderMatching,LHuillier1992CalculationsNm}. Solid HHG, on the other hand, is understood to occur regardless of phase mismatch, and the breaking of this phase matching condition has been proposed as an explanation for why solid HHG emission efficiencies are lower than the high density of excited electrons would suggest \cite{Li2024Bloch-WaveSolids}. Engineering crystal properties to improve phase-matching in the extreme nonlinear optics regime could therefore improve HHG efficiency in solids, however such improvements necessarily require precise understanding of the propagation of the harmonic waveform through the bulk material. 

The significance of macroscopic effects including pulse propagation in HHG in solids has been recognized over the past several years both in theoretical innovation \cite{Ghimire2012GenerationCrystals,Floss2018iAbSolids, Floss2019IncorporatingTheory, Orlando2019MacroscopicSemiconductors, Kilen2020PropagationGeneration, Rudenko2022Maxwell-SemiconductorSlabs, Kolesik2024PropagationSamples} and experimental investigation of surface versus bulk contributions in both trivial \cite{Hussain2021SpectralEffects, Korobenko2023InSolid} and topological materials \cite{Kim2022TheoryPicture}. As many thin and 2D materials can be challenging to work with under high fields due to their mechanical fragility and low damage thresholds \cite{Natoli2002Laser-inducedForms,Tsibidis2022DamageThickness}, HHG in bulk solids offers a more facile platform for applications like tabletop attosecond XUV pulse sources for ultrafast spectroscopy. Harmonics generated in the transmission geometry are also often more intense and easier to manipulate within the constraints of a laboratory relative to harmonics generated from the reflection geometry, although the intensity of reflected harmonics will vary with the optical properties of the specific crystals. 

It has been theorized that much of harmonic generation in the transmission geometry occurs at the back-surface of the crystal as higher order harmonics from visible to mid-IR driving lasers often fall beyond the transmission window of common solids, from semiconductors like silicon to insulators like sapphire, and thus are often reabsorbed or reflected within the bulk \cite{Yamada2023PropagationFilms}. Yet, even in this case, one must be able to simulate propagation of the broadband field pulse in order to determine its profile at the crystal back surface. Understanding propagation is also important for extracting dynamical information about materials and linking output spectra to underlying electronic structure. For example, accounting for propagation-induced chirp could help distinguish between intra- and interband harmonic generation mechanisms \cite{Ghimire2019High-harmonicSolids}.

We therefore require mechanistic understanding across micro- and macro- scopic length scales to facilitate the manipulation of attosecond pulse generation resulting from propagation through bulk solids and thereby realize the potential of HHG as a practical tool in applications including attosecond spectroscopy. In particular, we seek a computational framework which is both simple enough to provide physical intuition about which microscopic processes contribute to the output spectra yet retains sufficient complexity to capture non-ideal factors inherent in experimental system. \textit{Ab-initio} electronic structure methods to describe HHG in solids have been used to capture the microscopic response \cite{Floss2018iAbSolids,Tancogne-Dejean2017ImpactSolids,Tancogne-Dejean2017EllipticityDynamics} and have recently been employed in combination with propagation equations \cite{Floss2019IncorporatingTheory}. While such numerical methods have been laudably successful in replicating experimental observations, density functional theory (DFT) -based approaches are often computationally expensive and opaque to intuitive physical interpretation. Semiclassical models, on the other hand, provide a straightforward interpretation of the microscopic mechanism \cite{Parks2020WannierSemiconductors,Osika2017Wannier-BlochSolids,Yue2021IntroductionTutorial,Ortmann2021High-harmonicSolids} due in part to the analogy with the three-step model of HHG in gases \cite{Lewenstein1994TheoryFields, Corkum1993PlasmaIonization, Thorpe2025ASolids}. Yet, numerically simulating HHG in solids using semiclassical equations in conjunction with wave propagation remains sparsely studied, despite increased interest in the problem over the last several years, due to its conceptual and computational complexity \cite{Kolesik2024PropagationSamplesb, Kilen2020PropagationGeneration,Rudenko2022Maxwell-SemiconductorSlabs,Hussain2021SpectralEffects,Orlando2019MacroscopicSemiconductors}. We therefore present a theoretical framework which allows estimation of wave propagation effects when the light-matter interaction strength beyond the perturbative regime, in order to facilitate the interpretation of experimental HHG spectra so as to provide physical intuition about the underlying process.

In this paper, we report the implementation of the semiconductor Bloch equations (SBEs) to model the nonlinear polarization during wave propagation, where the SBEs simulate the nonlinear polarization which generates the nonlinear current term in Maxwell's equations, acting in place of the susceptibility tensors conventionally used to describe the nonlinear response in perturbative nonlinear optics. In contrast to some prior methods which simplify Maxwell's equations to the Helmholtz wave equation for the electric field \cite{Wu2022MultiscalePulses}, we opt to solve the full vectorial Maxwell's equations for the electric and magnetic components of the electric field in the time domain. Using the SBEs extends the capabilities of this propagation framework to describe light-matter interactions of very intense (on the order of $10^{12}\text{ W/cm}^2$) fields with bulk solids. We also discuss the significance of modeling the refractive index to describe the linear material response corresponding to transitions to electronic states not explicitly treated in the SBEs, and discuss options for its implementation. 
As a demonstration of the capabilities of this framework, we present simulations of HHG from a few-cycle near-infrared driving field in thin slabs of crystalline silicon; we present transmission and reflection spectra for several thicknesses; and lastly note that we may use this framework to compare the full Maxwell's equations approach with the unidirectional wave equation method of HHG propagation also implemented in this framework.

\section{Methods Overview}
In this section we first review the length-gauge SBE scheme for describing the interaction of the classical electric field with a two-band system representing the electronic structure of a crystalline solid. We describe the electronic structure calculations, in this case via density functional theory, used to obtain electronic band dispersion and transition dipole magnitudes for material parameters in the SBEs. Next, we describe the methods of numerical implementation of wave propagation, with a focus on the integration of the full vectorial Maxwell's equations. To describe propagation of the microscopic HHG response, we implement wave propagation in the Lightwave Explorer software originally designed for pulse propagation of perturbative nonlinear optics. This software is optimized for numerical efficiency and user accessibility, enabling precise and intuitive control over input pulse characteristics including group delay dispersion, pump-probe delay, and incident angle. One other powerful feature of this software is that it is written in a multilingual format with a combination of Compute Unified Device Architecture (CUDA) C and C++ using hardware-adaptive parallelization schemes to maximize computational efficiency across multicore central processing units and/or graphical processing units (GPU) \cite{Karpowicz2023OpenSimulation}. 

The overall scheme of the framework is as follow: in the initialization step, the input pulse beam profile is constructed in the frequency domain (e.g., Section 3.2) and Fourier transformed into the time domain to represent the field as a function of time ($t$) and transverse coordinate ($x$) at the initial $z$-position along the propagation axis, where the pulse parameters (central frequency, bandwidth, energy, spectral phase, carrier envelope phase, beamwaist) are provided in a run-time specified input text file. The material band dispersions, transition dipole moments, and dispersion derivatives are calculated within the framework from analytical functions fit in advance of numerical computations, derived in this work from density functional theory electronic structure calculations as discussed below, for the compile-time specified number of crystal momentum points and crystal lattice constant. 

The field is then advanced in time according to Maxwell's equations. The field evolves first through a $1.0 \mu m$ vacuum buffer before interacting with the material properties and, after leaving the material, evolves through another $1.0 \mu m$ back buffer. In our modified code, the nonlinear current response of the medium within Maxwell's equations is calculated from numerical integration of the SBEs. This is added to the current response calculated from the series of Lorentz oscillators that represent contributions from an empirically determined refractive index. This calculation of the material response occurs at every timestep alongside calculation of the field throughout its evolution. 

To calculate the material response, we first choose band dispersions and transition dipole moments to be used as material parameters in the SBEs, which we elaborate on below; we also choose refractive index resonances to be represented as Lorentz oscillators as a best fit to experimental data as discussed at the end of Section II.C. After the propagation has completed, the fields are saved as a function of space and time as well as Fourier transformed and integrated over the beam area to provide output emission spectra. Each element of this framework is expanded upon in greater detail in the following sections. 

\subsection{\label{sec:level2}Microscopic response}
Several methods exist to describe HHG in solids, including solving the time-dependent Schrodinger equation \cite{Wang2018InterferenceSolids,Li2018LimitationsGeneration} and ab-initio time-dependent density functional theory methods \cite{Floss2018iAbSolids,Floss2019IncorporatingTheory,Tancogne-Dejean2017ImpactSolids,Tancogne-Dejean2017EllipticityDynamics}. One common method to describe high harmonic generation (HHG) in solids is numerically solving semiconductor Bloch equations (SBEs) \cite{Luu2016High-orderApproach, Yue2021IntroductionTutorial, Golde2008HighExcitations, Vampa2015All-OpticalStructure, Vampa2015SemiclassicalCrystals, Kilen2020PropagationGeneration, Hagen2021ProbingSemiconductors, Pfeiffer2020IterationPropagation, Ortmann2021High-harmonicSolids, Han2019ExtractionGeneration, Gu2022Full-Brillouin-zoneMedia, Kolesik2023AssessmentMaterials, Kolesik2023NumericalSolidsb}. The microscopic response, in the context of solid HHG in this work, refers to the isolated emitted current generated from the SBEs in the absence of propagation. SBEs consist of a set of coupled partial differential equations that describe the time evolution of the density matrix representing the state of a periodic solid in response to a classical electric field under the dipole approximation \cite{Kira2011SemiconductorOptics}. From the density matrix formalism, off-diagonal elements of the density matrix can be mapped to the microscopic polarization ($p_k^{he}$, Equation 1) between electrons $e$ and holes $h$ in pairs of bands and the diagonal elements can be mapped to the carrier densities($f_k^e,f_k^h$; Equation 2,3) within each band \cite{Yue2021IntroductionTutorial}. 

\textit{SBE implementation}: Two gauge freedoms exist within these equations: a choice of structure gauge, relating to the representation of the electronic structure of the material through the band dispersion and transition dipole moment matrix elements; and a choice of laser gauge, relating to the representation of the electric field \cite{Yue2020StructureSolids}. In principle, the results of solving the SBEs should be equivalent regardless of the choice of gauge, yet in practice this is not the case because numerical simulations require a finite basis. Much work has been dedicated to analyzing the benefits and limitations of each gauge choice within the context of simulating HHG with the SBEs \cite{Yue2020StructureSolids, Kim2022TheoryPicture}.
The choice of laser gauge, in particular, alters the form and convergence criteria of the SBEs. The length gauge couples the equations of motion for the carrier densities and interband polarizations together in reciprocal space, whereas the velocity gauge decouples the crystal momenta and therefore has a numerical advantage. However, the inclusion of the longitudinal dephasing in the velocity-gauge representation is less straightforward, and a rigorous treatment requires converting back to the length-gauge in between time steps in order to apply the longitudinal dephasing, thus increasing the numerical cost. In addition, including many higher-lying conduction bands is required to achieve convergence in the velocity gauge. The length-gauge implementation, in contrast, includes dephasing within the microscopic polarization equation of motion and requires less bands to achieve convergence \cite{Yue2020StructureSolids}. Other multiscale nonlinear optics frameworks have chosen to use the velocity-gauge \cite{Wu2022MultiscalePulses} in part to simplify the computation of the SBEs by reducing the system of $N_k$ coupled partial differential differential equations to a system of $N_k$ ordinary differential equations, however the dephasing parameters $T_2$ has had such a prominent role in the discussion of numerical HHG simulation literature over the past several years that we prefer to use a gauge with straightforward and rigorous treatment of this parameter to better place our work in the broader context of HHG simulation literature. Therefore, we express the SBEs in the the length-gauge formulation and restrict our model to two bands \cite{Yue2021IntroductionTutorial, Luu2016High-orderApproach}, with the initial conditions of the interband microscopic polarizations and intraband carrier densities set to zero. 

\begin{eqnarray}
i \frac{\partial}{\partial t} p_k^{he} = &&(\varepsilon_k^{e}+\varepsilon_k^{h}-\frac{i}{T_2}) p_k^{he}\nonumber\\
&& - (1 - f_k^{e} - f_k^{h})d_k^{eh}E(t)\nonumber\\
&&+ iE(t)\nabla_k p_k^{he}
\end{eqnarray}
\begin{eqnarray}
\frac{\partial}{\partial t}f_k^e = \frac{\partial}{\partial t}f_k^e|_{relax} - 2Im[d_k^{eh}E(t)(p_k^{he})^*]\nonumber\\
+ E(t)\nabla_kf_k^e 
\end{eqnarray}
\begin{eqnarray}
\frac{\partial}{\partial t}f_k^h = \frac{\partial}{\partial t}f_k^h|_{relax} - 2Im[d_k^{eh}E(t)(p_k^{he})^*]\nonumber\\
+E(t)\nabla_k f_k^h
\end{eqnarray}

It should also be noted that there exists some debate on the subject of physically appropriate longitudinal dephasing times ($T_2$) used in the SBEs \cite{Gu2022Full-Brillouin-zoneMedia, Kolesik2023NumericalSolidsb, Kilen2020PropagationGeneration, Yue2021IntroductionTutorial}. $T_2$ ultimately acts as a phenomenological damping constant which numerically reduces the oscillations in the calculated miscroscopic interband polarization remaining after the pulse interacts with the material, thereby reducing noise in the simulated spectra and producing more distinct harmonic peaks. We discuss our treatment of this this in more detail in Appendix B. 

The SBEs account for material properties through terms relating to the dispersions of the electron $\varepsilon_k^e$ and hole $\varepsilon_k^h$ bands, as well as the transition dipole moment amplitude $d_k^{eh}$ between the bands. We discuss the retrieval of these properties next.  

\textit{Electronic structure calculations}:  Material properties, including band dispersions and transition dipole moments, are calculated using density functional theory (DFT) implemented in the Vienna Ab-initio Simulation Package (VASP) \cite{Kresse1994iAbGermanium, Kresse1996EfficiencySet}. In this study, we choose our material to be silicon in the cubic diamond crystal structure ($F\bar{d}3m$) along its $\Gamma \text{X}$ high symmetry path using the projector augmented wave (PAW) silicon pseudopotential and using the Perdew-Burke-Ernzerhoff (PBE) parametrization of the exchange-correlation functional, with a plane-wave cutoff energy of 540 eV and Gaussian smearing with $\sigma=0.1$ with a 11 x 11 x 11 Monkhorst k-point mesh for the self-consistent field calculation and 40 k-points between high symmetry points for the non-self-consistent field band structure calculation in line-mode. We do not include phases of the transition dipole moments in our SBE calculations, which is an acceptable approximation for centrosymmetric, topologically trivial materials \cite{Jiang2020SmoothSolids}, however we do account for the crystal momentum dependence of the transition dipole moment magnitude as calculated with the post-processing software \texttt{vaspkit} \cite{Wang2021VASPKIT:Code}. 

We calculate band dispersions and transition dipole moment amplitudes along 1D paths through the Brillouin zone between high symmetry points, and our 1D path approximation is justified by assuming a linearly polarized laser interacting with a single axis of symmetry in the crystal. Although there is some discussion of the validity of this assumption in the literature \cite{Gu2022Full-Brillouin-zoneMedia}, it is a widely used approach in the treatment of HHG \cite{Vampa2015LinkingSolids, Yue2021IntroductionTutorial}. As discussed in the propagation methodology section, we may account for the laser interacting with off-axis crystal directions by solving the SBEs along additional high-symmetry paths corresponding to off-axis electric field components. Note that silicon has six-fold symmetry in the $\Gamma \text{X}$ direction, which means all orthogonal high symmetry axes to $\Gamma \text{X}$ have identical electronic structure.

\begin{figure}[hbt!]
    \centering
    \subfigure[]{
    \includegraphics[width=1.0\linewidth]{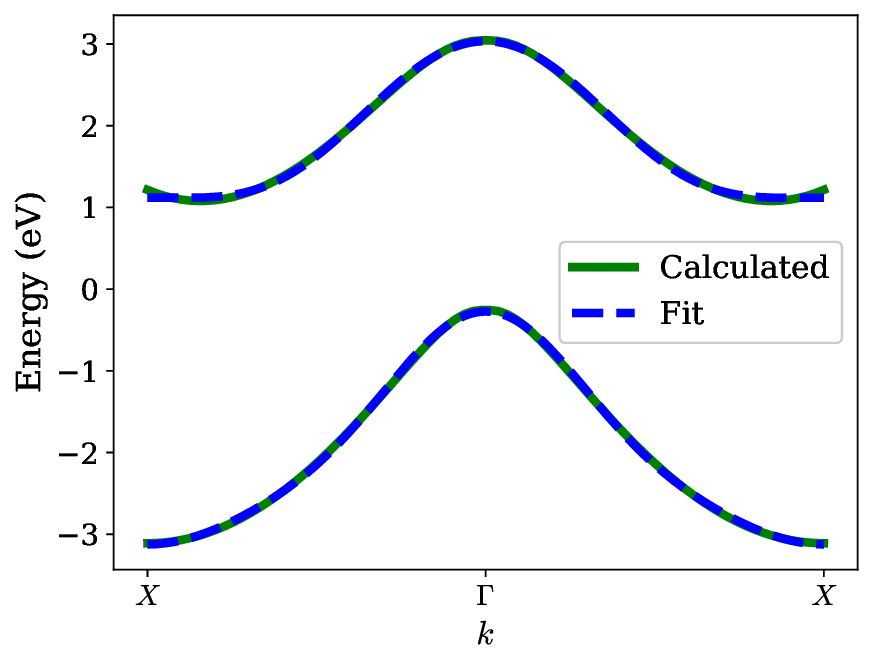}}
    \subfigure[] {\includegraphics[width=1.0\linewidth]{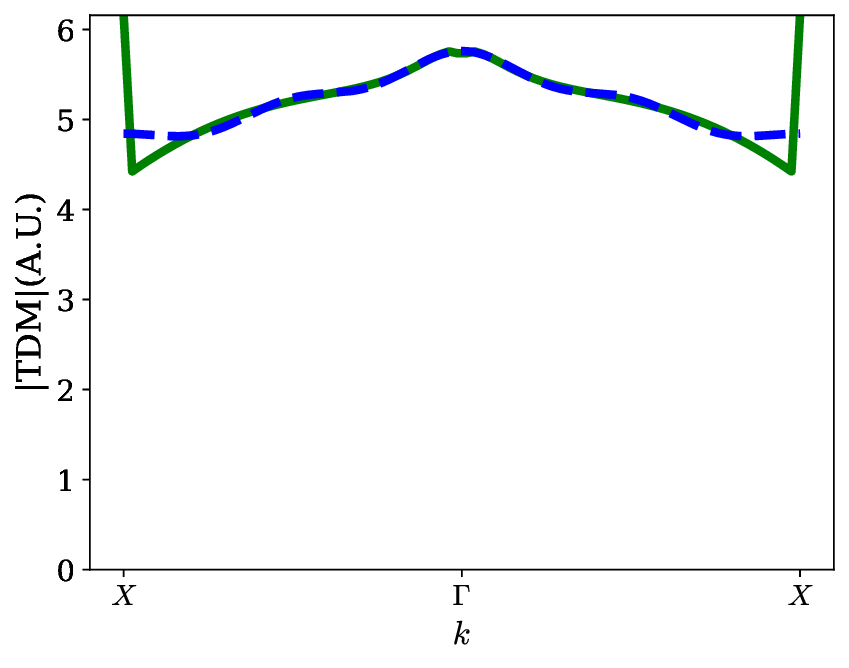}}
    \caption{Band dispersions (a) and transition dipole moments (TDM) (b) calculated in VASP and fit to sum of cosines for $\Gamma \text{X}$ high symmetry path in cubic-diamond silicon, with reciprocal crystal momenta in atomic units (A.U.). The TDM boundaries, which rapidly change character at the $\text{X}$ point, are neglected by the model since we find that smooth and continous derivatives are helpful to avoid generating numerical artifacts in the simulation which manifest as spurious high frequency components in the spectra.} 
    \label{fig:placeholder}
\end{figure}

We fit the band dispersions and transition dipole moment magnitudes to analytical functions composed of a sum of cosines in order to calculate the values at arbitrary $k$-points since many $k$-points may required to achieve convergence of the microscopic current calculated from the length-gauge SBEs \cite{Yue2020StructureSolids}, and then perform convergence tests on the SBE output by varying the number of $k$-points along the 1D path as shown in Figure 1. We find that 200 $k$-points along the $\Gamma \text{X}$ path are sufficient to achieve convergence of the SBE microscopic current in the case of our chosen material properties in silicon, and this relatively low number of $k$-points is likely a consequence of low curvature of the bands and TDMs in this particular material.  

Band dispersions for the highest lying valence and lowest lying conduction bands along the $\Gamma X$ path in cubic diamond silicon as shown in Figure 1 were fit to a sum of cosines for each band $\nu$ of the form:

\begin{equation}
E_k^\nu = \sum_n a_n \cos{(nk\frac{2 \pi}{L})}
\end{equation}

Where $L$ is the length of the Brillouin zone path and $n$ is the index corresponding to the element of the sum, and $k$ is the reciprocal space coordinate along a 1D path. The transition dipole moment magnitude between the valence and conduction band was also fit to a sum of cosines. The coefficients used in this study are tabulated below. 
\begin{center}
\captionof{table}{Coefficients from fitting Equation 4 to DFT calculated dispersions for the conduction band $a^c_n$, valence band $a^v_n$, and transition dipole moment matrix element magnitudes $a^t_n$. The $n=0$ terms in the cosine sum correspond to constant.}
\begin{tabular}{||c c c c||} 
 \hline
 $n$ & $a^c_n$ & $a^v_n$ & $a^t_n$\\ [0.5ex] 
 \hline\hline
 0 & 1.85276 & -1.92365 & 5.19805 \\
 1 & 0.97439 & 1.36351 & 0.41717\\ 
 \hline
 2 & 0.22400 & 0.22474 & 0.03021\\
 \hline
 3 & -0.01675& 0.06152 & 0.04151 \\
 \hline
 4 & - & - & 0.07399 \\
 \hline
 \hline
\end{tabular}
\end{center}

Care was taken to ensure smoothness and periodicity of the function and its first and second derivative at the boundaries of the Brillouin zone for each fit. The TDM boundaries in particular, which rapidly change character at the X point, are neglected by the cosine model fit because we found that material parameterizations used in the length-gauge SBEs (which contain gradients with respect to the crystal momentum; see Equations 1,2,3) require smooth and continuous derivatives everywhere and especially at the periodic boundaries in order to avoid generating numerical artifacts in the simulation which manifest as spurious high frequency components in the spectra. This is important because high frequency artifacts result in excessive numerical dispersion when subsequently propagated through the wave equation, leading to divergence and termination of the simulation.

\subsection{Pulse Propagation Frameworks}

Most pulse propagation frameworks capture nonlinearities described by the second- and third- order nonlinear susceptibility tensors, including second harmonic generation, third harmonic generation, self-phase modulation, and the optical Kerr effect. These susceptibility tensors are sufficient for describing perturbative nonlinearities usually arising from multiphoton excitation \cite{Boyd2020NonlinearOptics}. We extend this framework to the extreme nonlinear optics regime \cite{Wegener2005ExtremeOptics} in order to describe HHG by including the SBEs as a source of the nonlinear material response as an alternative to the susceptibility tenors. Within our framework we have several options for treatment of the wave propagation: the unidirectional pulse propagation equation (UPPE) mode which includes 2D cartesian, 2D radial symmetry, and 3D cartesian modes; and the time-domain Maxwell's equations mode which includes 2D and 3D cartesian propagation modes. Here, we focus on the propagation by solving Maxwell's equations in the time domain, and later compare with the UPPE propagation method.

In comparison with other studies of HHG propagation \cite{Kilen2020PropagationGeneration, Rudenko2022Maxwell-SemiconductorSlabs, Wu2022MultiscalePulses}, our framework provides both two- and three- dimensional descriptions of beam propagation which include the full spatial characteristics of the beam including specifying beam-waist at the focus and subsequent divergence as well as an option to upload an experimentally obtained frequency resolved optical grating (FROG) spectra of the pulse as the simulation input. These features enable one to compare theoretical results more precisely with experiment and account for the effects of these additional parameters.

\textit{Propagation via Maxwell's Equations}
Coupling the time domain solution to Maxwell's equations with semiclassical or ab-initio models of the microscopic HHG response is one approach studying HHG in bulk solids \cite{Yamada2023PropagationFilms,Rudenko2022Maxwell-SemiconductorSlabs,Floss2018iAbSolids}. 
This method captures forward and backward propagation effects which may be important in thin materials. Because this method advances the fields in the time domain, it can be fully parallelized over all three spatial dimensions, which is beneficial for HHG propagation simulations in which resolution of the z-step in the propagation direction must be small enough to resolve high frequency harmonics ($dz \leq \lambda_{min}/10$) but in which the time duration need not be too long since the pulse is so short (in practice, HHG driving laser pulses must often be short enough to achieve high field amplitudes without damaging the material). However, this time-domain solution method is especially sensitive to spatial and temporal step size, which are restricted by the Courant-Friedrichs-Lewy (CFL) criteria for convergence (see Appendix A), and also its associated numerical cost scales poorly with system size so it is only feasible to treat thin samples \cite{Taflove2005ComputationalMethod}.  

We couple the SBEs to the Maxwell's equation evolution of the electromagnetic field which solve in the time domain (Equation 5, 6) using a fourth order Runge-Kutta (RK4) method with sixth order in error approximations to the first order time and space derivatives of the electric and magnetic components of the fields.

\begin{equation}
\nabla \times \mathbf{E} = -\frac{\partial \mathbf{B}}{\partial t} 
\end{equation}
\begin{equation}
\nabla \times \mathbf{B} = \mu_0 \left( \mathbf{J} + \epsilon_0 \frac{\partial \mathbf{E}}{\partial t} \right)
\end{equation}

The use of the RK4 method prevents unconditional stability and thus minimizes the possibility of producing physically nonsensical output signal in the guise of a converged result. 

Within each RK4 step, the current $J$ in equation 6 is derived in part from the total current output from numerical integration of the SBEs as a system of coupled partial differential equations with respect to crystal momentum and time using the fourth order Runge Kutta (RK4) method stepping through time simultaneously with the time propagation of the Maxwell's equations. The crystal reciprocal space momentum derivatives within the SBEs approximated to only second order as this was found to provide optimal numerical stability while maintaining a minimum number of reciprocal momentum points required to achieve convergence. Note that the SBE output includes both linear and nonlinear contributions to the field evolution. While the propagation according to Maxwell's equations is calculated in the beam coordinates, the field is transformed into the crystal coordinates before each calculation of the current derived the SBE response, and the SBE response is then transformed back to the beam coordinate frame before being inserted back into Maxwell's equations as the source current as calculated from Equations 7-9. 
\begin{equation}
\mathbf{J}_{\text{inter}} = (\frac{2\pi}{a_0 N_k})^2\frac{\partial}{\partial t} \sum_k d_k^{eh} \, \text{Re}[p_k^{eh}]
\end{equation}
\begin{equation}
\mathbf{J}_{\text{intra}} = -(\frac{2\pi}{a_0 N_k})^2 \sum_{k, \lambda} \left( \nabla_k \, \varepsilon_k^{\lambda} \right) f_k^{\lambda}
\end{equation}
\begin{equation}
\mathbf{J}_{\text{total}} = \mathbf{J}_\text{inter} + \mathbf{J}_\text{intra}
\end{equation}

\textit{Refractive index treatment in time-domain Maxwell's equation}: The number of bands which can be included in the length-gauge SBEs is limited by computational efficiency \cite{Yue2020StructureSolids}, and while a two-band model is a decent approximation to HHG in centrosymmetric solids \cite{Jiang2018RoleCrystals}, such a model is likely insufficient to capture couplings to higher-lying conduction bands \cite{Hawkins2015EffectDielectrics}. Using the refractive index of the material to describe transitions to higher-lying bands is one way of accounting for this discrepancy.  

In the Maxwell's equations propagation mode, the refractive index is treated numerically as a system of ordinary differential equations representing a series of Lorentz oscillators (Equation 10, 11)  contributing the current $J$ driving the magnetic field evolution (Equation 6) and thus in turn to the electric field evolution (Equation 5); therefore, the refractive index when modeled as a series of oscillators will obey causality and thus maintain the self-consistency of this solution to Maxwell's equations. These Lorentz oscillators are advanced in time alongside the electric and magnetic fields, similarly via the RK4 method. The purely Lorentzian form of the refractive index may also be used in the UPPE mode, although the implementation uses the frequency-domain steady-state solution to the Lorentz oscillator since the wave equation in the UPPE mode is solved in the spectral domain as discussed below and later demonstrated in Section 3.7.

\begin{equation}
\frac{\partial \mathbf{J}}{\partial t} = k_L a_n \mathbf{E} - \gamma \mathbf{J} - \omega_0^2 \mathbf{P}
\end{equation}
\begin{equation}
\frac{\partial \mathbf{P}}{\partial t} = \mathbf{J}
\end{equation}

The coefficients $k_L$, $\gamma$, and $\omega_0^2$ model the coefficients of the refractive index represented by the steady-state solution to the coupled oscillator equations (Equation 10, 11) given by Equation 12:
\begin{equation}
n(\omega) = a_0 + k_L\sum_n \frac{a_n}{\omega_0^2 - \omega^2 + i\gamma \omega}
\end{equation}

In particular, $k_L = \frac{e^2}{\epsilon_0 m_e}$ where $m_e$ is the (effective) mass of the electron,$e$ is the electron charge, and $\epsilon_0$ is the vacuum permittivity; $a_0$ is a constant offset; and $a_n$ is the numerator of the $n^{th}$ Lorentzian oscillator. This equation can be fit to experimentally or computationally derived refractive index data as a function of frequency to obtain the appropriate coefficients of the oscillator. The resonance $\omega_0$ of each term physically corresponds to an absorption band of the crystal.

In our Lorentzian fit to the refractive index, we neglect the full height and width of the resonance found at the frequency of the material minimum direct band gap because the SBEs already describe electronic absorption at that energy (Figure 2). Higher frequency resonances in the refractive index that might represent excitation to higher lying bands are included as higher frequency resonances in the sum of Lorentzians as detailed below. Refractive index data was obtained from \cite{Schinke2015UncertaintySilicon,Shkondin2017Large-scaleMetamaterials} and smoothed and splined to remove the resonance corresponding to a direct band gap of 3.3 eV to avoid double-counting the linear shaping of the fundamental frequency by the SBEs. The resulting data was fit to a sum of seven Lorentzian terms of the form of Equation 12 using the \texttt{sellmeierFit} function from the LightwaveExplorer python module. The coefficients used for the simulations presented in this work are given in Table 2.

\begin{center}
\captionof{table}{Coefficients of the Lorentzian refractive index model}
\begin{tabular}{||c c c c||} 
 \hline
 $n$ & $a_n$ & $f_0 (THz)$ & $\gamma (THz)$ \\ [0.5ex] 
 \hline\hline
 0 & 1.00017  & --- & --- \\ 
 \hline
 1 & -4.730e+30  & 3.470e+02 & 1.016e+18 \\ 
 \hline
 2 & -1.045e+26 & 2.560e+01 & 9.422e+15 \\
 \hline
 3 &  4.618e+28 & 2.462e+07 & 8.219e+16 \\
 \hline
 4 &  1.399e+29 & 1.018e+03  & 1.745e+03 \\
 \hline
 5 &  1.399e+09 & 1.0181e+03 & 1.745e+03  \\ 
 \hline
 6 & -2.456e+28  & 3.000e+03 & 3.799e+17 \\ 
 \hline
 7 & -3.875e+29 & 6.600e+03 & 7.962e+17 \\ 
 [1ex] 
 \hline
\end{tabular}
\end{center}

\begin{figure}[hbt!]
    \centering
    \includegraphics[width=1.0\linewidth]{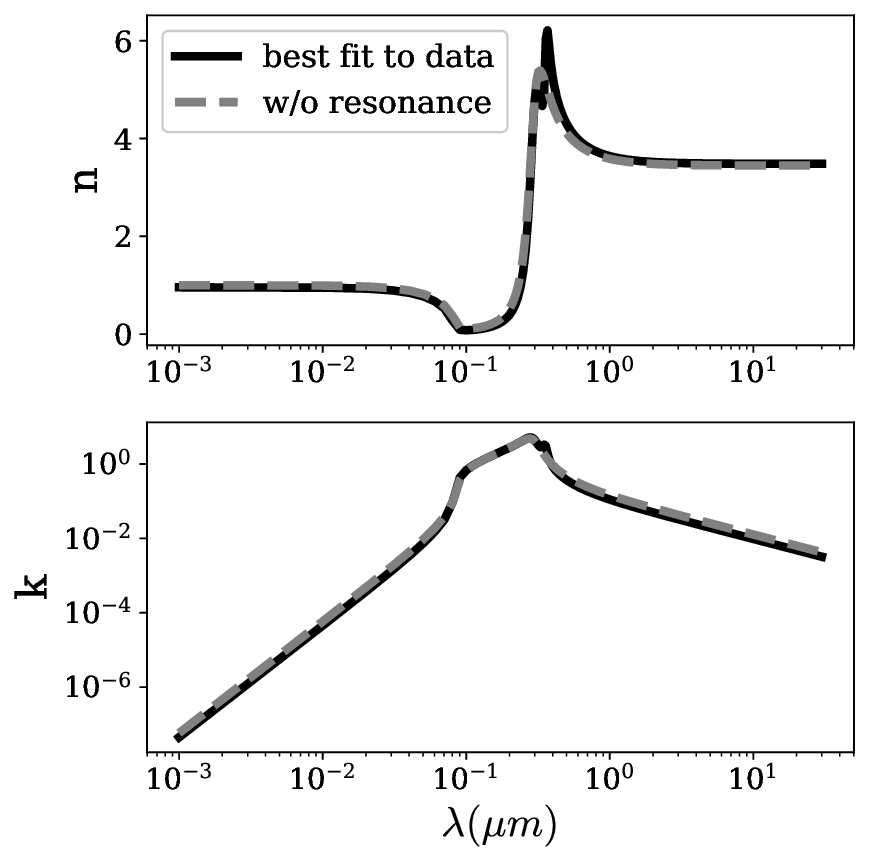}
   \caption{real (n) and complex (k) components of the Lorentzian refractive index model for silicon, with the solid black line showing a model including a small resonance near the minimum direct bandgap and the dashed grey line excluding the direct bandgap resonance as the SBEs should account for transitions resonant with the direct bandgap described by the conduction and valence band dispersions.} 
    \label{fig:placeholder}
\end{figure}

\textit{Collection of reflected or transmitted emission signal}: Both the electric and magnetic components of the field are evolved in the time domain solution of Maxwell's equations, but to minimize memory usage, only the electric field, sampled at a specific observation point along the propagation axis, is saved to the output array. The two tunable parameters in this recording process relevant to the collection of reflection versus transmission geometry spectra are the \texttt{waitFrames} and \texttt{observationPoint} variables. 

The grid is sampled beginning at a time set by the \texttt{waitFrames} variable which dictates how long the program waits before it begins saving the field to the output array. We discuss the construction of this more precisely in Appendix A.

The \texttt{observationPoint} corresponds to the point along the propagation axis at which the electric field values along the transverse direction are recorded. For the reflection geometry simulations, we set the observation point equal to a value within the micron front buffer prior to the start of the material. Conversely, for transmission geometry simulations, we set the observation point equal to a value within the back buffer after the end of the material. Finally, at the end of the time propagation, vectorial correction are made to the near field spectra to produce the the final time-signal and spectrum.

Wave propagation simulations over two or three dimensions which require fine temporal and spatial resolution are often inhibited by large memory allocations required to store all of the variables and fields calculated at each space-time mesh coordinate. The memory allocation in this code is designed to minimize such computational expense. The 5.0 $\mu m$ propagation depth simulations performed in this study parallelized across only 16 CPUs at 98$\%$ efficiency required approximately 5.5 hours and 10 GB of memory. Message passing interface (MPI) parallelization is used to distribute the cores in the time domain Maxwell's equation mode over all spatial coordinates so that the time stepping can be sequential. When using CPU-based parallelization, the number of cores should be a multiple of eight in order to maximize utilization efficiency. 

\textit{Unidirectional Pulse Propagation Equation}
The unidirectional pulse propagation equation (UPPE) is an approximation to the wave equation which sufficiently describes many nonlinear optical processes while remaining numerically tractable in part due to its neglect of backward propagating components of the wave \cite{Couairon2011PractitionersSimulation}. Variations on the unidirectional wave equation have been used to simulate pulse propagation within the context of filamentation \cite{Andreasen2013MidinfraredMethods}, few-cycle pulse propagation \cite{Brabec1997NonlinearRegime,Brabec2000IntenseOptics}, and high harmonic generation in gases \cite{Brabec2000IntenseOptics,LHuillier1992CalculationsNm}. The form of the wave equation used here is carrier-resolved and thus sufficient for describing few-cycle pulses \cite{Brabec1997NonlinearRegime}, which is a requirement for HHG processes in which short pulse durations are often required in order to reach pulse amplitudes high enough to reach the tunneling ionization regime without inducing material degradation.
\begin{eqnarray}
 \frac{\partial{}}{\partial z} \tilde{\textbf{E}}(\textbf{x},\omega) =&i \sqrt{k^2-k_\perp^2}  \tilde{\textbf{E}}(\textbf{x},\omega)\nonumber\\
& + \frac{i\omega ^2}{2\epsilon c^2 \sqrt{k^2-k_\perp^2}} \textbf{P}^{NL} (\textbf{x},\omega)
\end{eqnarray}
The UPPE propagation of HHG in solids, however, has been explored only recently in a few instances \cite{Kilen2020PropagationGeneration, Kolesik2004NonlinearEquations}. In \cite{Kilen2020PropagationGeneration} an implementation of the UPPE considering both $k_x$ and $k_y$ momenta was coupled with 1D HHG from length-gauge SBEs to describe the even- and odd- order harmonic response from a GaAs crystal from a 10.6 micron driving laser in order to investigate the interplay of propagation and ultrafast dephasing ($T_2$) often required to produce distinct high harmonic peaks in the plateau region of the output spectra from numerical simulations. The authors cite a 600um propagation simulation with resolution up to the 40th harmonic (which enforces an upper bound on the maximum step size in the propagation direction) as requiring around 8 days of computational time, reinforcing the numerical cost of incorporating propagation effects in HHG simulations. 

In the UPPE mode of our framework, the SBEs are coupled to a carrier-resolved non-paraxial unidirectional propagation equation \cite{Karpowicz2023OpenSimulation}. One should here be careful to note that the UPPE is in some literature only used to refer to the paraxial version of the unidirectional wave equation, although here we use the term more generally. To derive the UPPE, we assume a wave equation expressed in the frequency domain, and consider only forward propagation of the electric field and thus neglect the backward propagator. Within this the slowly evolving wave approximation (SEWA) which assumes the field amplitude and phase evolve sufficiently slowly along the propagation coordinate relative to the central wavelength \cite{Brabec1997NonlinearRegime, Kolesik2012QuantifyingPropagation}. So, we can neglect the second-order derivative of the field with respect to the propagation coordinate, and arrive at a unidirectional carrier-resolved wave equation (Equation 13). We maintain non-paraxiality in the implementation of this equation through the expansion $K(\omega) = \sqrt{K^2 - K_\perp}$, which provides a more accurate and more numerically tractable equation than its paraxial counterpart. The UPPE in this framework is solved using an RK4 scheme to advance the field in the frequency domain. The nonlinear current and polarization source terms are calculated in the time domain at each z-point estimation of the electric field. 

In this framework, we solve the two-band SBEs (Equation 1-3) as a system of partial differential equations using the RK4 scheme at each update of the electric field to obtain the microscopic polarizations and carrier densities at each point in the reciprocal crystal space and at each spatial grid point, then sum over the crystal momentum points of the unit cell to obtain the macroscopic interband (Equation 14) and intraband (Equation 15) current output from the SBE response. 

\begin{equation}
\mathbf{J}_{\text{inter}} = \frac{\partial}{\partial t} \sum_k d_k^{eh} \, p_k^{eh}
\end{equation}
\begin{equation}
\mathbf{J}_{\text{intra}} = -\sum_{k, \lambda} \left( \nabla_k \, \varepsilon_k^{\lambda} \right) f_k^{\lambda}
\end{equation}

The equations are then solved iteratively, similar conceptually to approaches described in prior literature \cite{Kilen2020PropagationGeneration}. When SBE mode is turned on by setting $T_2 \neq 0$, calculation of the perturbative nonlinear response from the empirical susceptibility tensors is entirely neglected to avoid double-counting of the nonlinear response. 

The UPPE mode allows four options for modeling the material refractive index. First, a generalized Sellmeier equation which models the real component of the refractive index and thus may be used to describe dispersion and diffraction of the electromagnetic field in the material \cite{Voronin2017TheAir}. While Sellmeier coefficients may be temperature dependent \cite{DeFranzo1993IndexWavelengths, Thomas1997FrequencySapphire}, we assume a room temperature dispersion for the purpose of these studies, although we note that the instantaneous thermal energy of the material may vary due to the impingement of high energy laser light. Second, a combination of the real Sellmeier terms and complex Lorentzian terms may be used to model the complex refractive index capturing dispersion, diffraction, and absorption of light by the material. The third option is a purely Lorentzian form of the refractive index which is in principle equivalent to the sum of Lorentz oscillators describing the refractive index in the time-domain Maxwell's equation propagation mode. Fourth, the complex refractive index can be modeled as a series of Gaussian resonances to more accurately describe the Urbach tails featured in the absorption spectra of solids \cite{Urbach1953TheSolids, Vadlamani2020Tunnel-FETShape}.

\section{Propagation of HHG in a semiconductor from few-cycle mid-IR driving laser pulse}
We present HHG spectra from silicon, an indirect bandgap semiconductor material with a direct minimum band gap $\simeq 3.3$ eV, from a 2.0 $\mu m$ driving laser with a bandwidth of 18.3 THz correspond to a 24 fs full-width half-maximum Gaussian pulse envelope (SI.4). We choose silicon because it is ubiquitous in the fields of electronics and optics, and has been widely studied in the context of HHG and associated strong field processes \cite{Suthar2022RoleSilicon, Yamada2023PropagationFilms, Klemke2019Polarization-state-resolvedSolids, Zimin2023DynamicPhotoinjection}.

First, we present the microscopic response, which is the emitted current calculated from the SBEs without any propagation. We then present the results of this system within our framework in the time-domain Maxwell equations propagation mode, and compare the spectra in both transmission and reflection geometries for a progression of propagation depths to observe how propagation may modify the harmonic spectra. We also discuss select numerical aspects related to the beam profile and low-frequency modes of the refractive index. Lastly, we compare with the UPPE mode of propagation.

\subsection{Microscopic HHG Emission}
First, we present the microscopic response, which is the emitted current calculated from the SBEs without any propagation. In the following section we present the results of this system within the propagation framework, and compare the spectra in both transmission and reflection geometries for a progression of propagation depths to observe how propagation may modify the harmonic spectra. We present here the results of the two-band SBE model for silicon along the $\Gamma \text{X}$ high symmetry axis, where spectra have been normalized to the amplitude of the driving laser fundamental frequency peak. In the spectra below, we use a dephasing time $T_2= 8$ \textit{fs}, which we discuss further in Appendix B.

\begin{figure}[hbt!]
    \centering
     \includegraphics[width=1.0\linewidth]{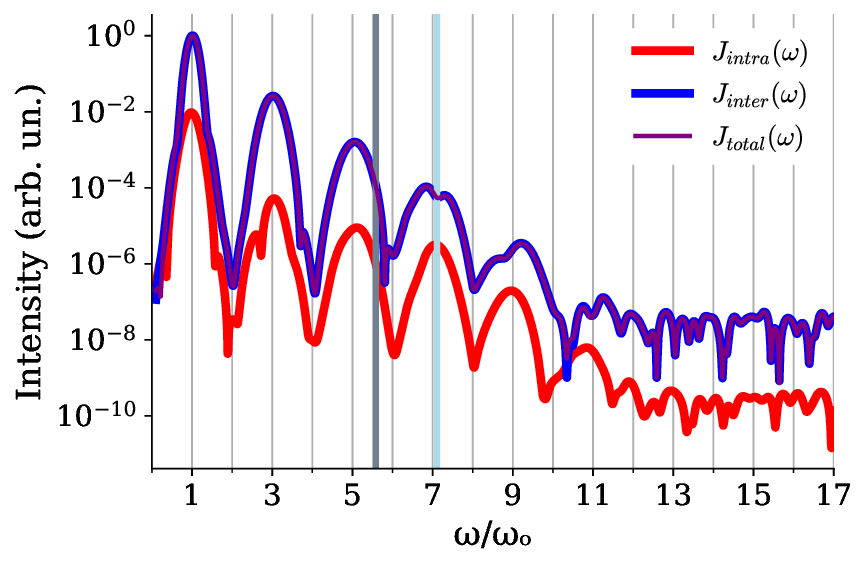}
    \caption{SBE response from driving field of central wavelength $\lambda=2.0 \mu m$, 24 \textit{fs} duration, $E_0=1.94\times10^9$ $V/m$ along $\Gamma \text{X}$ high symmetry axis in silicon, decomposed into intraband $J_{intra}(\omega)$ (red), interband $J_{inter}(\omega)$ (blue), and total current $J_{total}(\omega)$ (purple). The interband and total current contributions closely overlap, as the intraband current in this csae is several orders of magnitude smaller. The grey line represents the minimum direct band gap energy and the light blue line represents the maximum direct band gap energy.} 
    \label{fig:placeholder}
\end{figure}

The maximum direct energy gap described by the model band dispersion is equal to about seven times the photon energy of the $2 \mu m$ driving laser, denoted in the Figure 3 by the light blue vertical line, and indeed harmonic generation efficiency decreases after the seventh harmonic, and some spectral distortion is visible as expected from harmonics generated at band edges. Therefore we focus on only the first nine harmonics in the simulated propagated spectra for the majority of this work. 

\subsection{Input Electric Field}
We show the normalized on-axis time domain signal (Figure 4(a)) and 2D beam profile along transverse space and time dimensions (Figure 4(b)) for the input driving field. The equation used for construction of the electric field in the frequency domain is given by Equation 16:

\begin{equation}
\tilde{E}(\omega)=A_0e^{-(\omega - \omega_0)^{2}/\Delta_\omega^{2}}
\end{equation}
Where $\omega_0$ is the central frequency of the pulse, $\Delta_\omega$ is the bandwidth. We set the initial spectral phase of the pulse to zero and use Gaussian spectral shape, although higher order super-Gaussians can be used. Pulse energies specified in the input file were calculated to correspond to $\sim 10^{12} \text{ W/cm}^2$ intensity unless otherwise stated. Input parameters for the pulse were as follows:

\begin{center}
\captionof{table}{Input pulse parameters}
\begin{tabular}{||c c||} 
 \hline
 $\lambda_0$ & 2.00 $\mu m$ \\
 \hline
 $\Delta_\omega$ & 18.30 \textit    {THz}\\ 
 \hline
 $w_0$ & 5.00 $\mu m$\\
 \hline
 $I$ & 1.00 \textit{TW}$/cm^2$\\
 \hline
\end{tabular}
\end{center}

\begin{figure}[hbt!]
    \centering
    \subfigure[]{
\includegraphics[width=0.45\linewidth]{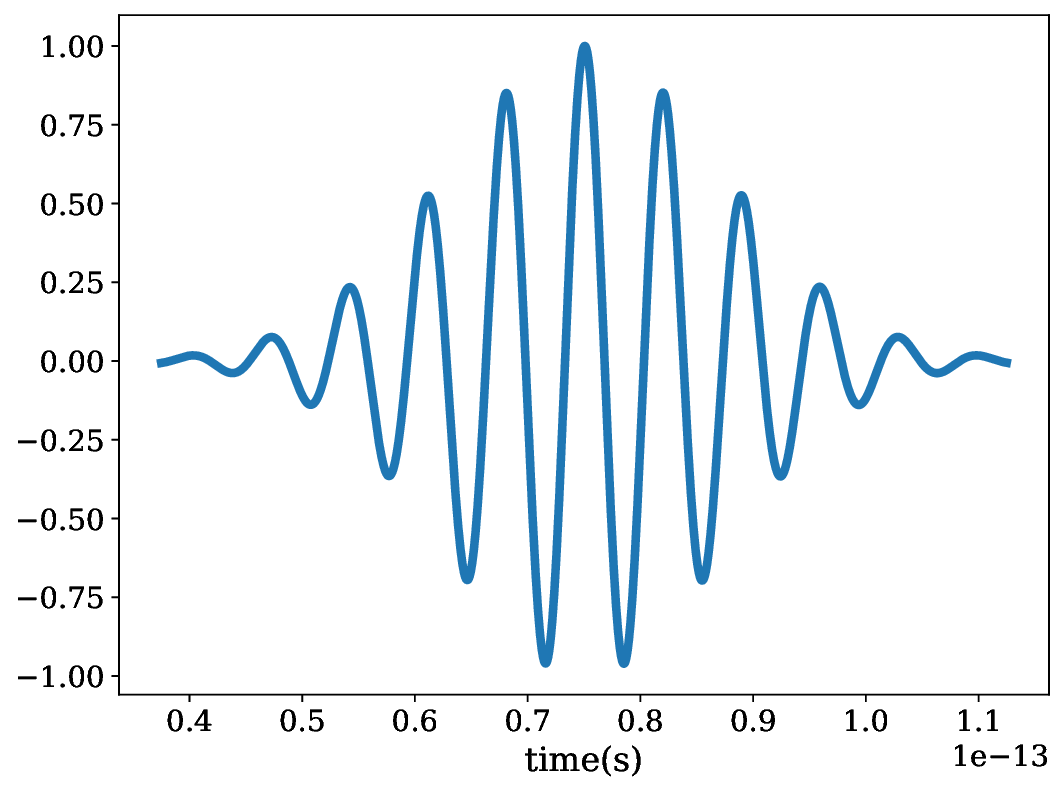} 
}
\subfigure[]{
\includegraphics[width=0.45\linewidth]{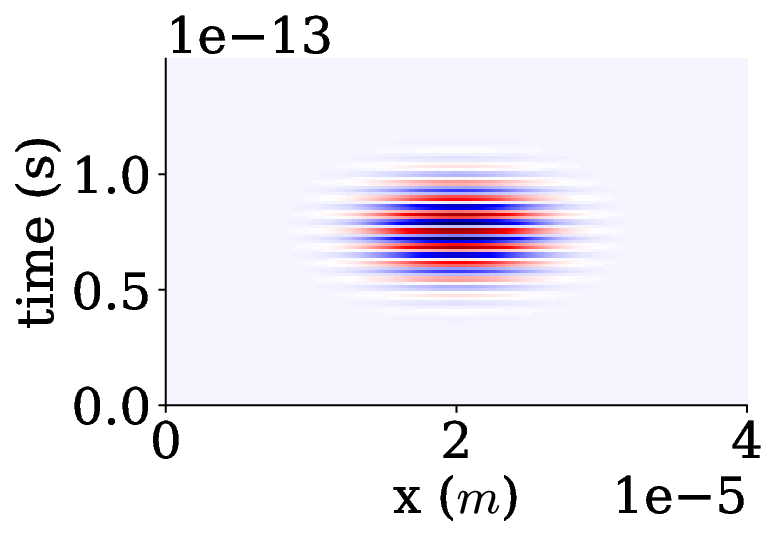}}
    \caption{Normalized input electric field (a) on-axis time signal (b) 2D electric field profile as function of time $t$ and transverse coordinate $x$ } 
    \label{fig:placeholder}
\end{figure}

\subsection{Low Intensity Driving Field Response}
Here in Figure 5 we include a comparison of the response from a low intensity ($10^{8} \text{ W/cm}^2$) driving field through thin film silicon from UPPE and time-domain Maxwell's equation propagation for depths of $z=50nm$ (Figure 5(a)) and $z=100nm$ (Figure 5(b)) to show that both propagation methods predict similar behavior in the linear propagation regime for these very short propagation distances. 
\begin{figure}[hbt!]
    \centering
\subfigure[]{
     \includegraphics[width=1.0\linewidth]{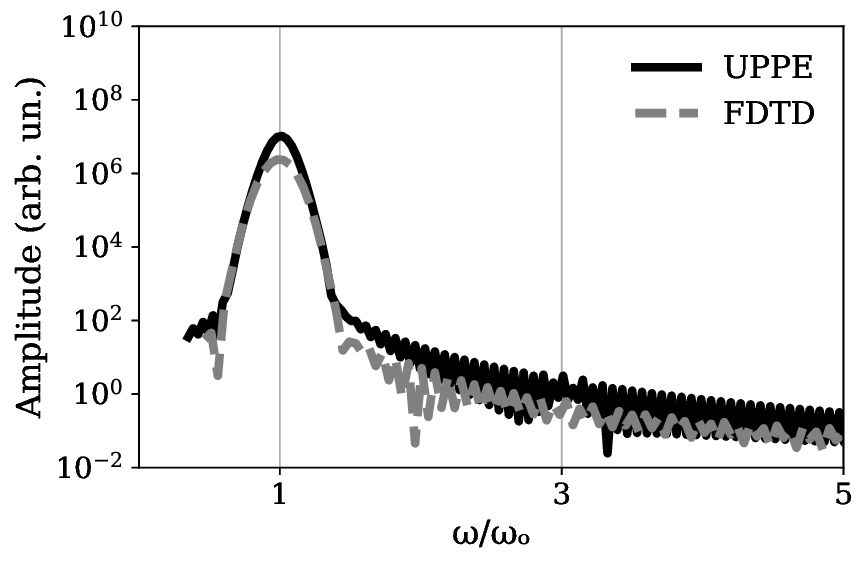}}
     \subfigure[]{   \includegraphics[width=1.0\linewidth]{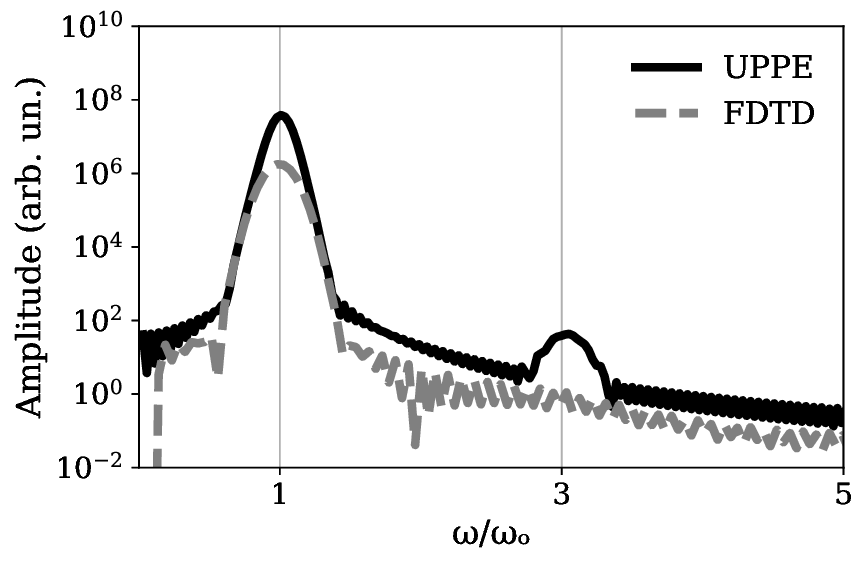}}
    \caption{Propagation of weak ($I = 10^{8}$ $\text{ W/cm}^2$) field centered at $\lambda = 2.0\mu m$ through (a) 50 nm and (b) 100 nm thin film silicon along $\Gamma X$ using UPPE mode versus using time-domain Maxwell's propagation mode (denoted here "FDTD" for brevity, although we note that our time-stepping algorithm differs from the FDTD convention).} 
    \label{fig:placeholder}
\end{figure}
The Maxwell's propagated spectra show a slightly lower intensity of the fundamental frequency peak because this propagation mode accounts for reflection as well as absorption, both of which detract from the transmitted light intensity, whereas UPPE only describes absorption since the wave propagation model neglects reflection. As expected, differences between the spectra are exacerbated by propagation, leading to higher on-axis intensity more quickly in the case of the UPPE propagation, leading to small amounts of third harmonic generation. We must note however that using the SBEs in their current form to describe weak light-matter interaction may contain inaccuracies as we employ the strong field approximation which allows us to neglect the many-body Coulomb term under the assumption that the light-matter interaction is much stronger than any Coulomb interactions in the material.

\subsection{HHG Propagation using Maxwell's Equations}
We compare propagation of the HHG response using the time-domain solution to Maxwell's equations for a series of short propagation distances from few nanometer to a few microns. We implement processing of the simulation outputs in python, including a Tukey window ($\alpha=0.5$) which is applied to the output time signal before applying the normalized fast Fourier transform using the Scipy FFT module. All spectra are calculated from on-axis ($x_{transverse}=0$) traces unless otherwise stated, as examined later in Section 3.5.
\begin{figure}[hbt!]
    \centering
    \subfigure[]{
    \includegraphics[width=1\linewidth]{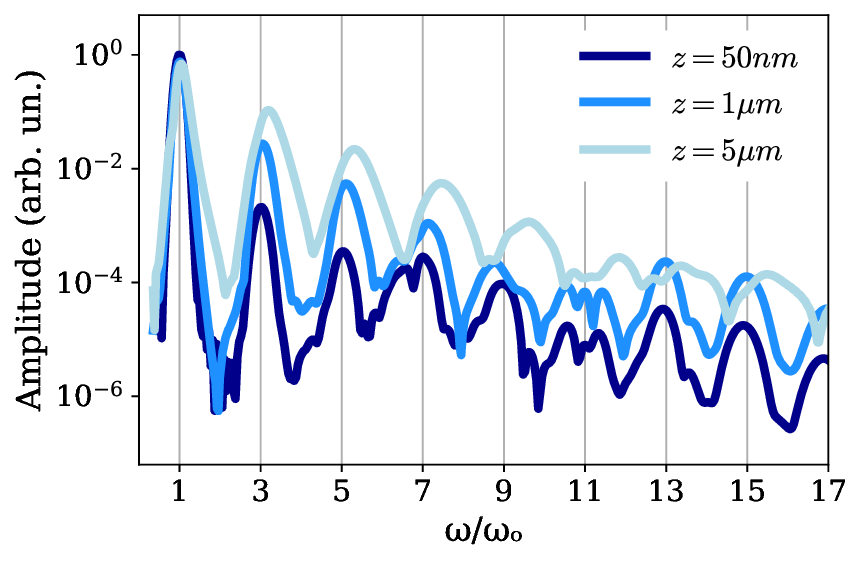}
    }
    \subfigure[]{ \includegraphics[width=1\linewidth]{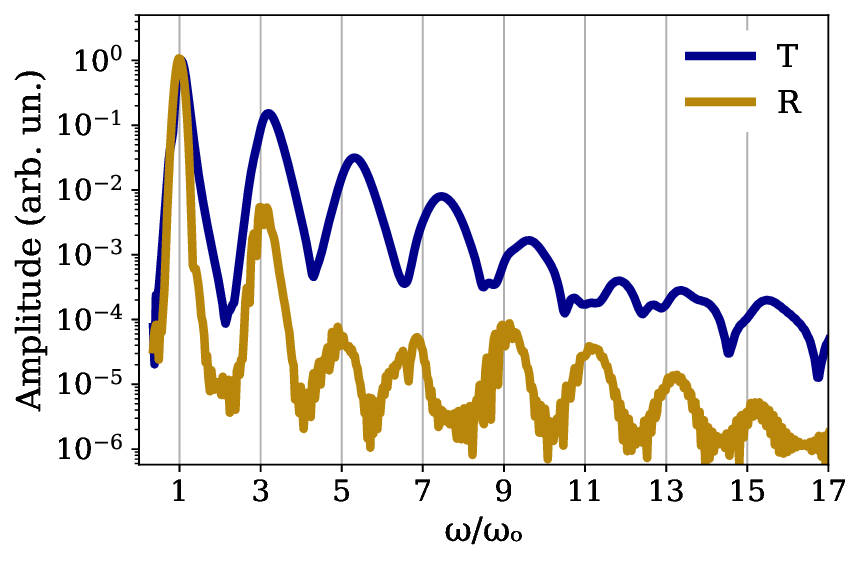}}
    \caption{(a) On-axis spectra from a series of short propagation distances from $z = 50 nm$ through $z = 5 \mu m$. (b) Comparison of transmission and reflection geometry simulated HHG spectra after $1 \mu m$ propagation distance through silicon.}
\end{figure}
Figure 6(a) demonstrates the modulation of harmonic intensity, lineshape, and central frequency with propagation distance. In particular, harmonics without propagation have narrower linewidths and are centered on the integer harmonic frequency. As the propagation distance is increased to 5 $\mu m$, the harmonic linewidths broaden and shift towards higher energies. We include in the figures the spectra at the fundamental frequency as well, so that we may note how the lineshape of the fundamental similarly broadens and develops a shoulder towards the blue. These results therefore suggest that it may be the propagation-induced reshaping of the fundamental component of the pulse which leads to similarly modulated harmonic spectra. 

The harmonic spectra in Figure 6(a) also show increased harmonic intensity as propagation distance increases, and a slight decrease in the peak intensity at the fundamental frequency. We note that the spectra presented here consist of on-axis traces of the electric field, so it would make sense if this intensity modulation is a result of increased on-axis beam intensity as a result of self-focusing effects. If the fundamental intensity has been enhanced by self-focusing by the time the light reaches the back surface of the crystal, we would indeed expect the generated harmonics to appear more intense. Although we find that the spectra corresponding to integration over the entire beam does not show significantly more differences in the harmonic intensities for different propagation distances relative to the on-axis trace (section 3.5), increased intensity near the center of the beam could indeed still lead to enhanced harmonic efficiency and thus an extended plateau region because harmonic generation scales nonlinearly with intensity. This increased harmonic intensity may also be attributed to slight accumulation of the nonlinear response in the case of harmonics below the bandgap or, for higher harmonics, an underestimation of high-frequency absorption described by model of the refractive index.

In Figure 6(b), we compare the harmonic spectra in reflection and transmission geometry from a thin crystal by moving the observation point at which the signal is monitored from after the crystal to before the crystal, as discussed earlier in Methods. The harmonic lineshapes in reflection are narrower and less blue-shifted than those of harmonics in the transmission geometry. The transmitted harmonic intensities are however higher than reflected harmonic intensities, which may in part be due to the depth-dependent harmonic intensity discussed for Figure 6(a), and is also largely dependent upon the amount of frequency-dependent dispersion of the refractive index. Namely, a strongly dispersive real component of the refractive index will lead to a more intense reflected signal overall. The conclusion is that the transmitted harmonic spectra are of higher intensity than the reflected harmonic spectra, which qualitatively aligns with previous numerical studies simulating harmonic propagation in nanometer-scale silicon thin films \cite{Wu2022MultiscalePulses}. 
\begin{figure}[hbt!]
    \centering
    \subfigure[]{
    \includegraphics[width=1\linewidth]{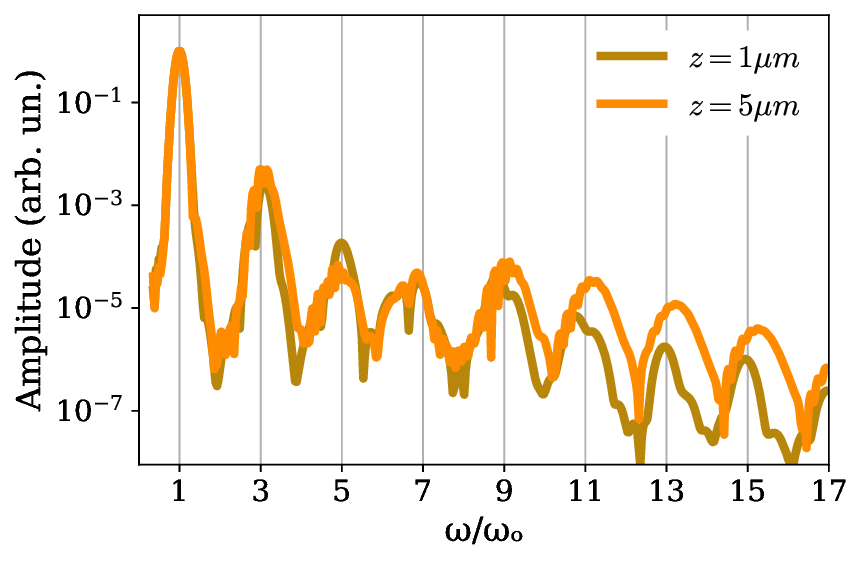}}
    \subfigure[]{
    \includegraphics[width=1\linewidth]{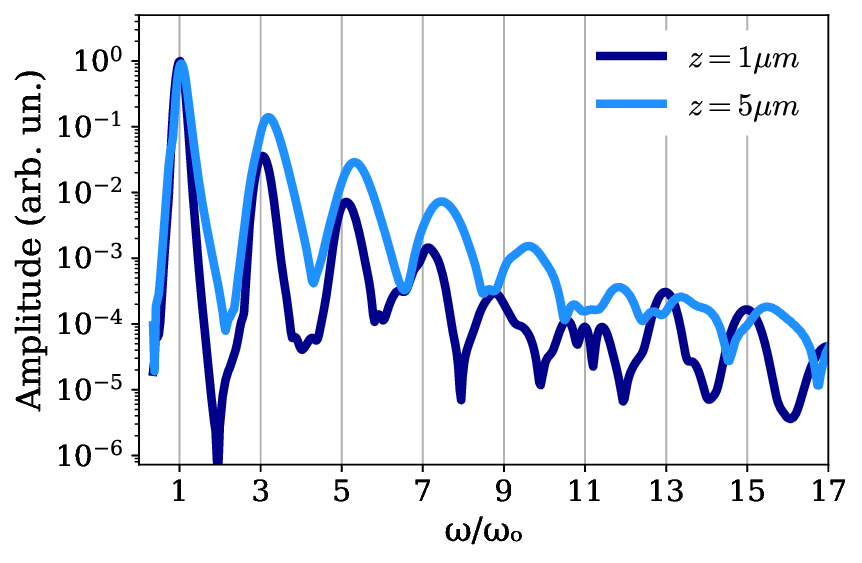}
    }
    \caption{On-axis spectra from two different propagation thicknesses ($z$=$1\mu m$ and $z$=$5\mu m$), comparing HHG from (a) reflection (observation point equal to 10 $z$-steps prior to crystal front surface) and (b) transmission (observation point equal to 10 $z$-steps after crystal back surface).}
\end{figure}
Figure 7 analyzes the observable effects of propagation in each detector geometry by comparing different propagation distances in both the reflected signal (Figure 7(a)) and the transmitted signal (Figure 7(b)). Reflected harmonics below the material direct bandgap energy are overall independent of material thickness as expected. This contrasts with transmitted harmonics below the bandgap energy which do show the thickness-dependent intensity and central frequency shifts clear in Figure 6(a) and Figure 7(b). 

The modulation of harmonics from propagation, as well as the relative intensities of reflected versus transmitted harmonics, are affected by the choice of refractive index, introduced in section 2.2. Including more resonances in the summation within the Lorentzian $n(\omega)$ (Equation 12) would allow one to further improve the fit to the experimental data. Also, these expressions for $n(\omega)$ clearly neglect many of the low-frequency absorption peaks since we found that absorption of frequencies much lower than the driving laser frequency do not significantly affect the simulation results (Section 3.6).

The simulated HHG spectra and observations presented here overall support the argument that propagation modulates the harmonic structure, and that even if the majority of observed harmonic yield may be generated at the back surface of the crystal, the propagation-induced reshaping of the fundamental still significantly affects the harmonic structure, and simulation of the reshaping of the field at the fundamental frequency still requires precise treatment of the strong higher order nonlinearities affecting its evolution. It is therefore crucial to account for propagation effects in solid HHG in order to make any concrete claims relating the transmitted HHG spectra to the underlying electronic structure of the material. In the following two subsections we provide more details on computational concerns relating to the above discussion.

\subsection{Propagated 2D beam profile analysis}
\begin{figure*}[hbt!]
\centering
\subfigure[]{
    \includegraphics[width=0.45\linewidth]{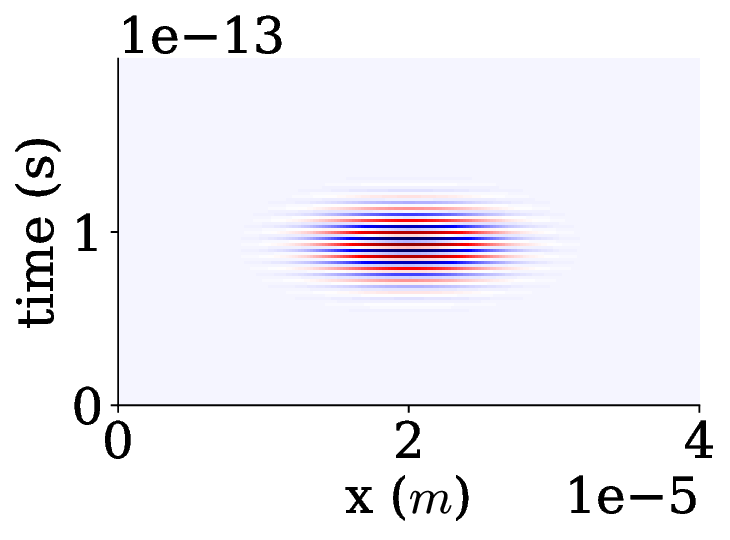}
    \label{fig:wide:a}}
\subfigure[]{
    \includegraphics[width=0.45\linewidth]{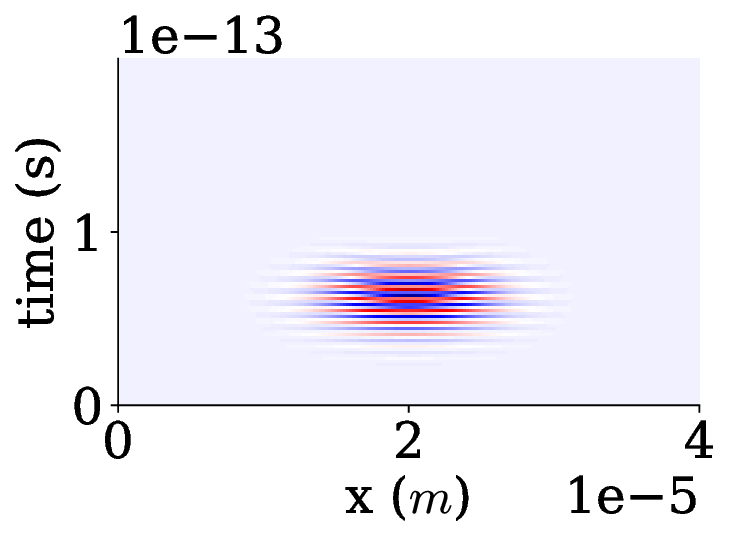}
    \label{fig:wide:b}}
\subfigure[]{
    \includegraphics[width=0.45\linewidth]{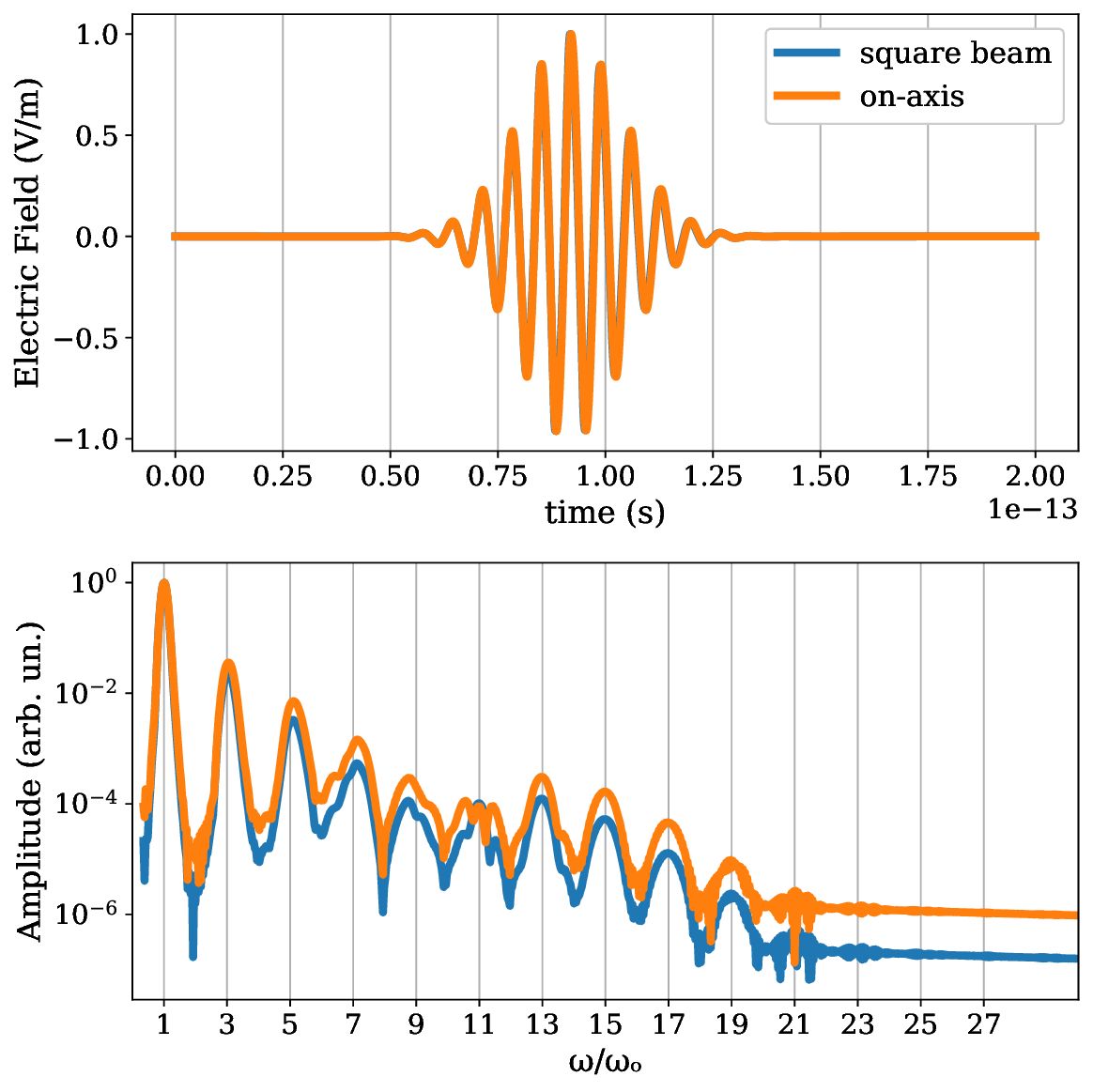}
    \label{fig:wide:c}}
\subfigure[]{
    \includegraphics[width=0.45\linewidth]{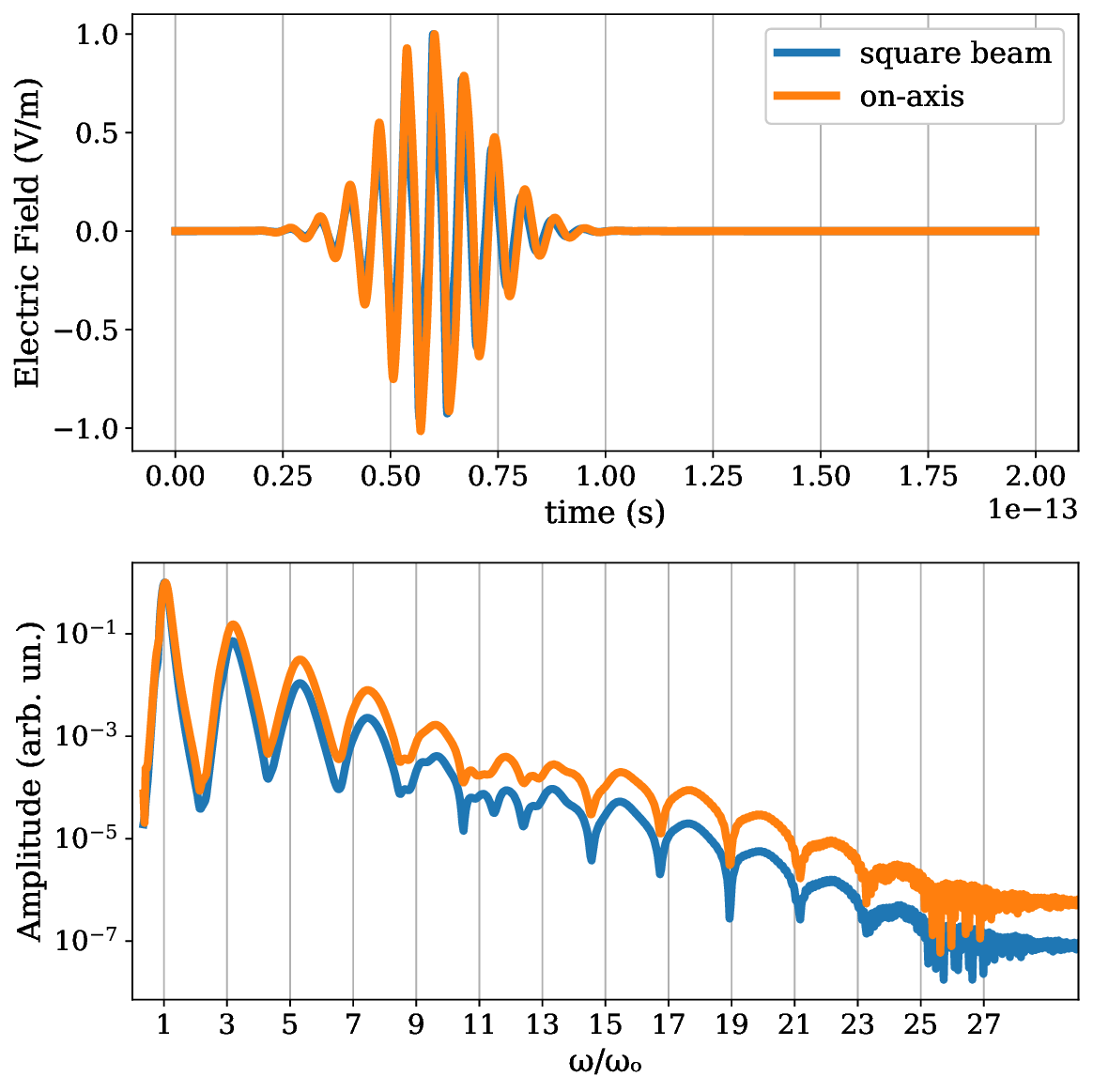}
    \label{fig:wide:d}}
    \caption{2D electric field profile for signal propagated through (a) $1 \mu m$ and (b) (a) $5 \mu m$ silicon as function of time $t$ and transverse coordinate $x$. Comparison of normalized, unfiltered on-axis signal (blue) versus the signal integrated with the entire area equally weighted (i.e. "square beam" approximation) (orange) for (c) $1 \mu m$ and (d) $5 \mu m$.} 
    \label{fig:wide}
\end{figure*}
We present in Figure 8(a,b) the spatial profile of the beam, as a function of time $t$ and transverse coordinate $x$, after propagation through (Figure 8(a,c)) $z=1\mu m$ versus through (Figure 8(b,d)) $z=5 \mu m$ of silicon. Also shown are the spectra after FFT of the unfiltered on-axis signal and "square beam" integrated signal with all $x$-points equally weighted in the sum. In the normalized spectra, the harmonic intensities of the on-axis traces are slightly higher, relative to the fundamental intensity, than in the case of the square beam. This is reasonable given that we would expect a higher proportion of the energy in the high intensity region of the beam to be converted to high harmonics compared with the lower intensity off-axis areas of the beam which naturally contribute less to harmonic generation. On-axis contributions therefore dominate the spectral features. The spatial profile of the beam after such short integration distances as $1\mu m$ shows minimal curvature of the wavefront. However, upon propagation to $5 \mu m$, some wavefront curvature begins to emerge. Therefore, for consistency, all the spectra presented in the main body of this work correspond to the on-axis trace. 

\subsection{Refractive index phonon modes}
In the framework as it has been presented thus far, indirect gap transitions cannot be explicitly represented numerically without including phonon motion in the model as our transition dipole moment integrals assume only contributions from points of equal crystal momenta. On one hand, it can be argued that the attosecond timescale of harmonic generation is much shorter than the typical picosecond timescale of phonon motion in a lattice, and so neglecting phonon motion may be a reasonable approximation. On the other hand, it is the interference of attosecond pulses generated over a finite pulse duration which shape the harmonic spectra, suggesting a longer significant timescale. Moreover, several recent studies have highlighted the coupling of phonon motion and strong field processes \cite{Hu2024Phonon-CoupledInteractions, Korolev2024UnveilingSolids, Hatch2026ProbingSpectroscopy}. We therefore mention two methods which could be implemented within our framework to account for phonons. One could compute a phonon-mediated band structure in VASP and compute microscopic polarization arising from interaction with that electronic structure in place of the static electronic structure calculations used here. Alternatively, one may include strong low-frequency oscillator modes in the sum of Lorentzian oscillators used to model the refractive index to approximate phonon modes in the material.  

Using the time-domain Maxwell's equations propagation we compare the propagated HHG spectra from a model without any low frequency resonances to the propagated HHG spectra from a model including one small low frequency resonance below the driving laser frequency, as depicted in Figure 9(a), and observe minimal difference in the output harmonics, at least for these short propagation distances of $5 \mu m$ shown in Figure 9(b). Longer propagation distances and stronger phonon absorptions may reveal more pronounced effects originating from such low frequency absorption features, and exploration of this is ongoing in our concurrent work. 

\begin{figure}[hbt!]
\centering
\subfigure[]{
    \includegraphics[width=1.0\linewidth]{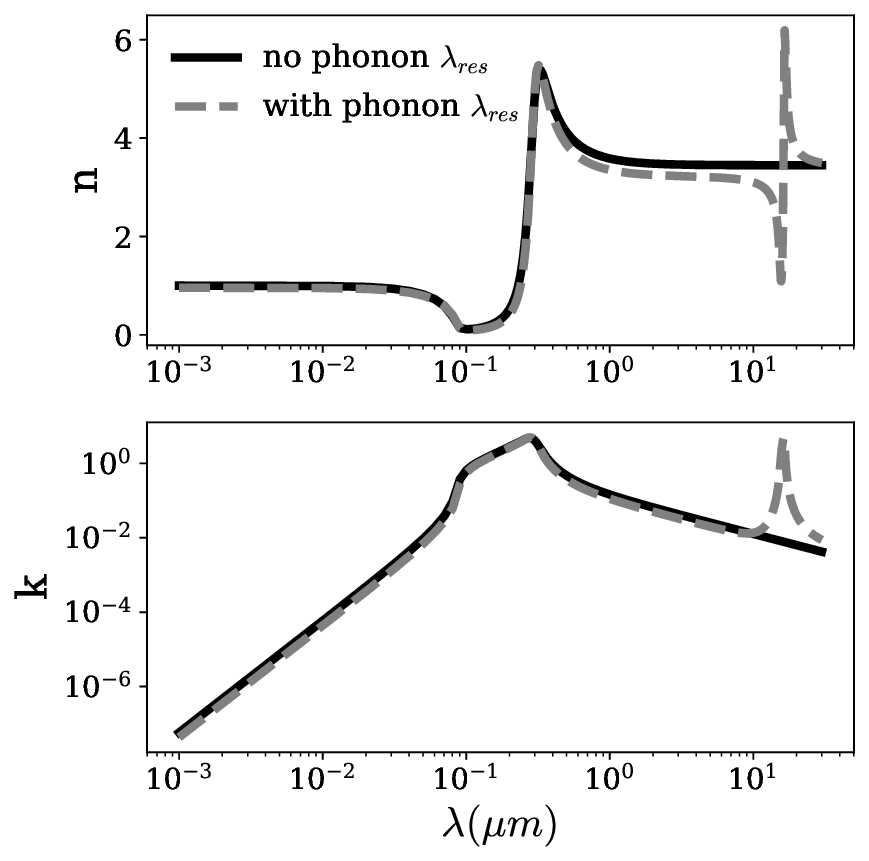} 
    }
    \subfigure[]{ \includegraphics[width=1.0\linewidth]{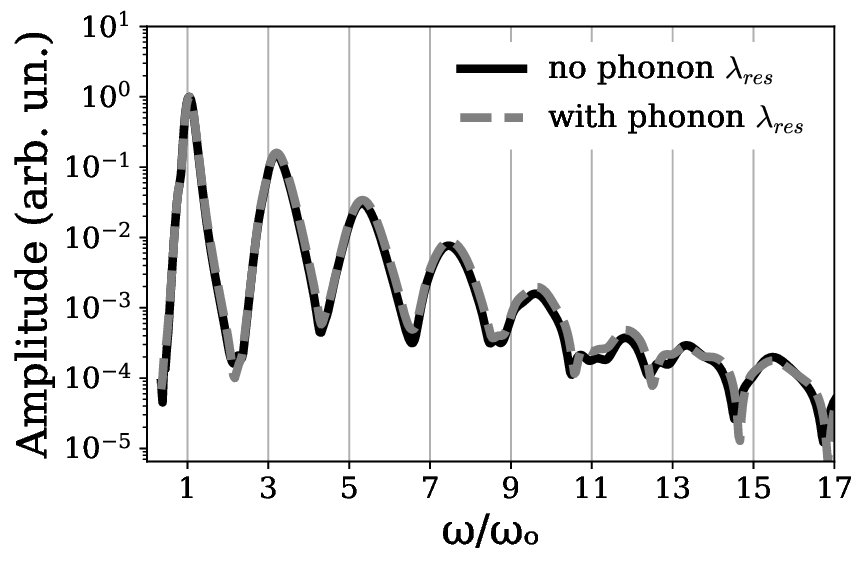}} 
    \caption{(a) Model of the refractive index without ("A") and with ("B") a low frequency phonon mode around $\lambda = 11 \mu m$; (b) Propagation of an intense driving field centered at $\lambda=2.0\mu m$ over $z$=$5\mu m$ compared for a refractive index model without versus with this low frequency "phonon" mode.}
    \label{fig:placeholder}
\end{figure}

\subsection{HHG propagation using the Unidirectional Pulse Propagation Equation in Si thin films}
The UPPE method, explained in section 2.2, is often a computationally cheaper alternative to full Maxwell's equations propagation, and this propagation model is also an option in the original LightwaveExplorer software. We therefore include a brief examination of solid HHG propagation through very thin slabs using the UPPE model in  silicon along the $\Gamma \text{X}$ axis using a Lorentzian refractive index model for several thin propagation distances below and above the driving laser wavelength. 

However, because the HHG process generates frequency components over an extremely broadband range, numerical dispersion and artifacts of asymmetric time signals in the spectral domain become a concern in this self-consistent spectral domain propagation. Propagating the electric field using the UPPE in the spectral domain while calculating the nonlinear material response in the time domain requires repeated application of the Fourier transform. Over many spatial propagation steps, small numerical errors accumulate to produce instability which either introduce inaccuracy via numerical artifacts or cause numerical integration to diverge. The time duration, temporal window function, and time step must therefore be optimized for each simulation and may depend on the propagation distance. Such concerns are particularly relevant in the context of HHG propagation for which the SBEs may produce a signal with small asymmetries in time, namely that the signal may be discontinuous at the boundaries of the time domain and thus not perfectly time-periodic as in principle inherently assumed by the Fourier transform algorithm. In the mode in which we solve Maxwell’s equations in the time domain (as discussed in section 2.2 and 3.4-3.3.6), this property of FFTs is does not affect the propagation of noise because the FFT is only applied at the beginning and at the end of the simulation. 

One way to address this is to apply a Tukey window after each calculation of the evolved electric field before the field is used in calculation of the nonlinear response in order to reduce numerical artifacts resulting from high-frequency components of the pulse traveling to the edges of the periodic time domain, specifically in cases where these high-frequency components are unphysical such as numerical artifacts from large higher-order derivatives of the dipole dispersion near the Brillouin zone edges. However, if such a window is implemented, care must be taken that minimal pulse energy is removed as a result of the windowing, although in the nonperturbative regime we expect harmonic intensities to vary sublinearly with driving laser intensity which should in principle reduce the effects of such numerical error. 

\begin{figure}[hbt!]
    \centering
    \includegraphics[width=1.0\linewidth]{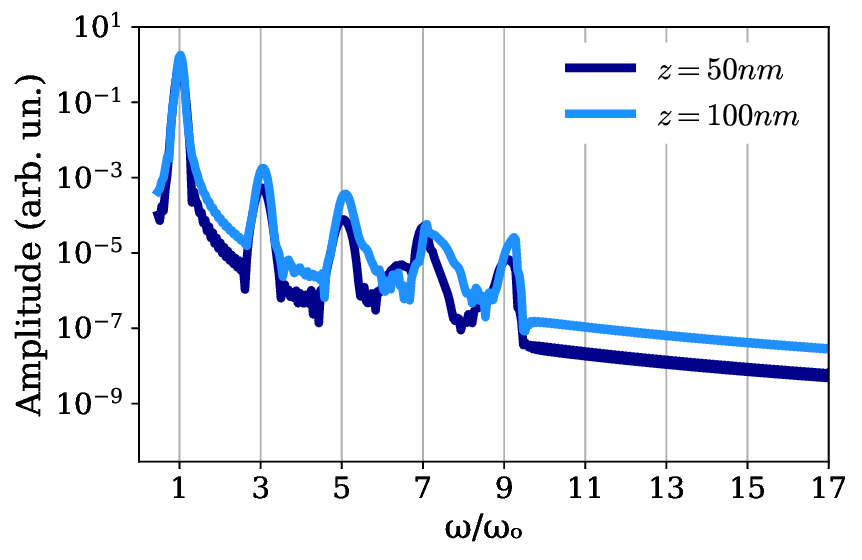}
    \caption{UPPE simulations from silicon propagated for z = 50 nm versus z = 100 nm.} 
    \label{fig:placeholder}
\end{figure}

We first examine UPPE models of HHG propagation in a series of thin silicon slabs where the refractive index is modeled as a series of Lorentzian oscillators for ease of comparison with the time-domain Maxwell's equations propagation approach. We can also compare the Maxwell's equations approach with UPPE wave propagation of harmonics over short distances. The low-order harmonic intensities, such as the third and fifth order harmonics, are of similar intensities when propagated in the UPPE and Maxwell's methods as expected. The quantitative agreement between the two propagation mode options within our single framework means that we can in principle use this software to compare how different numerical approaches to simulating propagation distort the harmonic spectra. 

\begin{figure}[hbt!]
    \centering
    \includegraphics[width=1.0\linewidth]{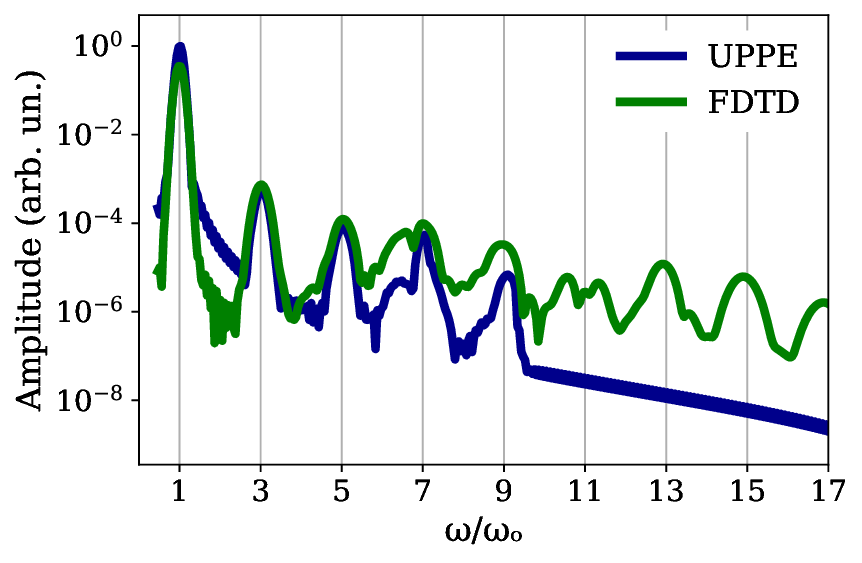}
    \caption{On-axis spectra for 50 nm comparing spectra calculated from the UPPE versus the time domain Maxwell ("FDTD") wave propagation methods. Discrepancies above the bandgap photon energy around 10$\omega_0$ are attributed to the numerical treatment of $n(\omega)$.}
\end{figure}

The spectra in Figures 10 and 11 also demonstrate the effects and limitations of the refractive index treatment in the two different propagation mode. In the UPPE mode, the refractive index is included in the electric field propagation kernel as a prefactor of the nonlinear current (Equation 13). In the propagation through Maxwell's equations, in contrast, each resonance in the refractive index is treated as a material oscillator. The UPPE spectra, in consequence, may show more prominent dependence on the refractive index resonances. In Figure 10, we note a sharp cut off in the signal after the ninth harmonic in the spectra of the UPPE propagated pulse, which, as shown in Figure 11, does not appear in the equivalent time-domain Maxwell's equation propagation. This is a result of a filter within the propagation code which determines that the value of the refractive index multiplied by the nonlinear current kernel is unphysical and sets the propagation kernel for these frequencies to zero. This feature may also show up prominently in cases in which the resonance corresponding to the bandgap frequency is erroneously included in the refractive model along with the SBEs which also describe transitions between the valence and conduction band. Such a “double-counting” of the valence-conduction band response tends to magnify the harmonic response at the bandgap frequency, which in UPPE may trigger a numerical correction to enforce stability and appear as a sharp increase or decrease in the spectral intensity around the resonance frequency. When using this HHG propagation model, it is therefore important to consider which resonances in the refractive index are already treated by the SBEs. This may vary if multiband SBEs are incorporated into the model.

Although we do neglect the narrow resonance corresponding to the direct band gap in our refractive index model, we leave in some amplitude in the real and imaginary components around the band gap region. So, to more definitively demonstrate that this sharp cutoff in the frequency spectra is a consequence of the refractive index treatment, we show in Figure 12 the results from an alternative model of the refractive index in which the bandgap resonance is entirely neglected in the refractive index model and instead only one sharp high frequency resonance along with a broad low frequency resonance are included in the model (Figure 13). We again set the driving wavelength to $2 \mu m$ but use slightly lower intensities and a shorter dephasing time $T_2=0.25 \tau_{\omega_0}$ to help with numerical convergence by reducing high frequency noise and thereby facilitate comparison of the propagation effects on shaping harmonic spectra.
\begin{figure}[hbt!]
    \centering
    \includegraphics[width=1\linewidth]{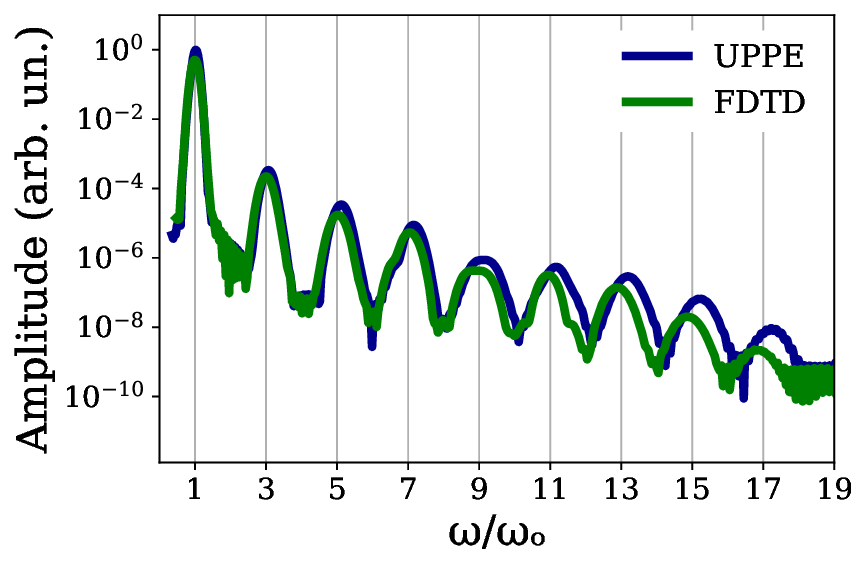}
    \caption{On-axis spectra for z = 50 nm propagation distance comparing spectra calculated from the UPPE versus the time-domain Maxwell equations "FDTD" wave propagation methods for a minimal model of the refractive index as depicted in Figure 13, with E=20 nJ and $T_2=1.7$ fs or $ 0.25\tau_{\omega_0}$.}
\end{figure}

With this difference in the refractive index, the sharp cutoff disappears and we are able to more easily compare the two different propagation modes. We do not expect the spectra to drastically differ for such short propagation distances so this functions more as an internal benchmark of the two propagation methods implemented in our framework. However, already the propagated spectra suggest that the UPPE mode may overestimate the harmonic intensities slightly in transmission as expected from a unidirectional picture which neglects the reflected component. 
\begin{figure}[hbt!]
    \centering
    \includegraphics[width=1\linewidth]{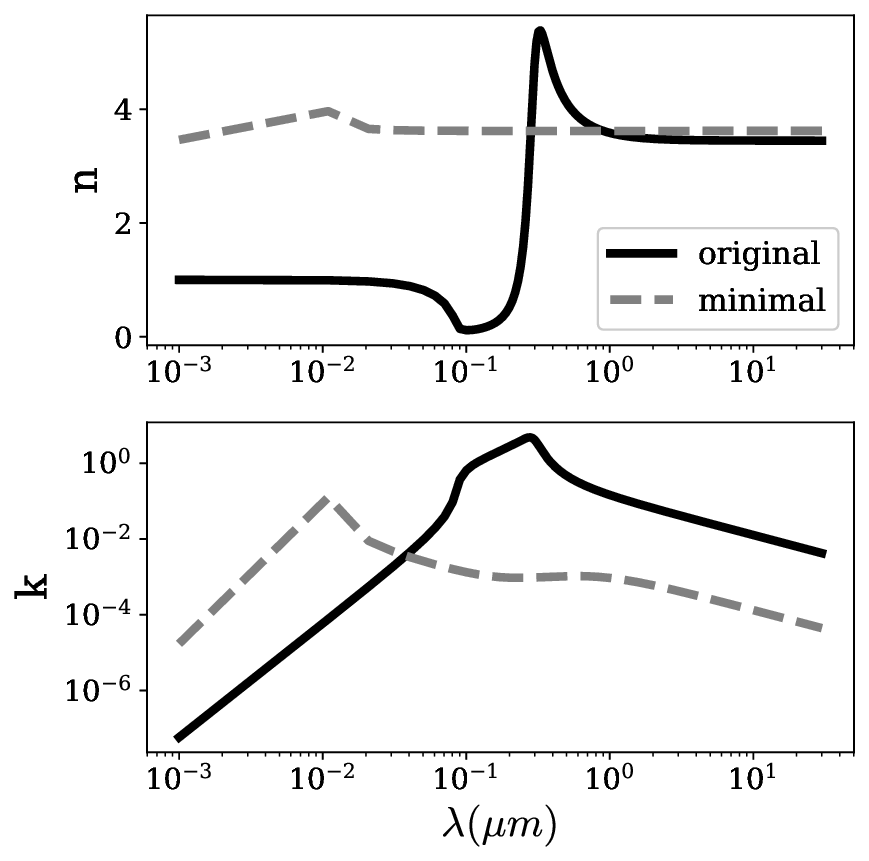}
    \caption{Plot of real (n) and imaginary (k) components of the refractive index model for silicon for the case of a minimal model including just one high frequency resonance and one lower frequency resonance used in the above simulations in Figure 12. ("minimal", dashed gray) versus the remainder of plots in this paper ("original", solid black).}
\end{figure}
\section{Conclusion}
We present a theoretical framework for coupling the non-perturbative nonlinear polarization from intense few-cycle light pulses with Maxwell's equations to describe the propagation of the response through bulk solids.  We integrate the SBEs to model the microscopic material response and couple this with a finite-difference solution to Maxwell's equations, with both systems of differential equations solved in the time-domain by a fourth order Runge-Kutta scheme. In this study, we demonstrate the use of this framework to study the HHG response of silicon thin films under illumination by strong, few-cycle mid-IR pulses. The results presented here emphasize how propagation modifies the microscopic HHG emission profile and thereby obscures the relation between the electronic structure of the material and the HHG spectra. Figure 6(a) and Figure 6(b) in particular highlight the modulation of the harmonic spectral lineshapes and intensities after propagation through the bulk. 

The development of a computationally efficient wave propagation framework which extends to the regime of extreme nonlinear optics therefore provides an accessible and physically intuitive model of HHG propagation through which future studies could systematically evaluate numerical approaches to simulating HHG in bulk solids under different material and driving laser parameters. Specifically, extensions of this project which are currently underway include comparing the propagated HHG spectra along different high symmetry axes in the same crystal; comparing HHG spectra from different crystals with varied bandgap energies and symmetries; and analyzing how features of the input field, including spectral dispersion, manifest in the output emission spectra. Such explorations contribute to our understanding of the interplay of macro- and microscopic mechanisms within different regimes of light-matter interaction, leading to enhanced utility of HHG spectroscopy as a probe of electronic properties and improved material engineering control of attosecond XUV pulses.

\section*{Acknowledgements}
A.N.H would like to thank Isabelle Tigges-Green and Jesse Griff-McMahon for helpful discussions during the conceptualization of this project, as well as Matthew Mason, Vedin Dewan, and Alfy Benny for helpful discussions during the writing of this manuscript. This work was partially supported by the Division of Chemical Sciences, Geosciences and Biosciences, Office of Basic Energy Sciences, of the US Department of Energy through grant no. DE-SC0015429; by the NSF under Grant No. PHY 2206711; and through the Princeton University’s Materials Research Science and Engineering Center DMR-2011750. A.N.H. gratefully acknowledges the support of the NSF through the Graduate Research Fellowship.

\section*{Author Contributions}
A.N. Hejazi: Writing -- original draft, Software, Methodology, Investigation, Formal analysis, Conceptualization. N. Karpowicz: Writing -- review \& editing, Software, Methodology. G.D. Scholes: Writing -- review \& editing, Supervision, Conceptualization, Funding acquisition. J.M. Mikhailova: Writing -- review \& editing, Supervision, Conceptualization, Methodology, Funding acquisition.
\bibliographystyle{elsarticle-num}
\bibliography{referencesWholeLibrary}

\newpage
\appendix
\section*{Appendix A: Runge Kutta Scheme for Time Domain Solution to Maxwell's Equations}
\setcounter{figure}{0}
\renewcommand{\thefigure}{A.\arabic{figure}}
\setcounter{equation}{0}
\renewcommand{\theequation}{A.\arabic{equation}}
In this appendix, we provide more explicit detail on the computational propagation scheme. The time-domain solution to Maxwell's equation implemented in Lightwave Explorer uses a fourth Order Runge Kutta (RK4) scheme to advance Maxwell's Equations in the time-domain. This is in contrast to the conventional Yee algorithm for a finite-difference time domain (FDTD) solution to Maxwell's equations \cite{Taflove2005ComputationalMethod}. 
The material oscillators corresponding to the SBEs and refractive index Lorentzians are advanced in time alongside the fields as in auxiliary differential equation approaches to FDTD propagation of electromagnetic fields in dispersive mediums \cite{Taflove2005ComputationalMethod}.
The spatial first-order derivatives of the electric and magnetic field components needed to calculate the curls of the respective fields on the lefthand side of Equations 5 and 6 of the main text are approximated using a sixth-order central differences scheme. The curl components are then included in the kernel expression for $\frac{\partial B}{\partial t}$ and $\frac{\partial E}{\partial t}$. The length-gauge SBEs consist of $N_k$ set of coupled partial differential equations, and the derivatives with respect to $k$ are computed with second-order accuracy central differences discretizations, as higher-order discretizations required finer $dk$ steps to reduce numerical instability, thereby significantly worsening computational cost. The SBEs are solved in atomic units. All input fields are converted to atomic units for the SBE computation and the output current is converted back to SI units for the field evolution. This is because the SBE computations would otherwise involve values which are too small and result in many numerical errors. 

The total number of time points $N_t$ over which the field is propagated is given by $N_t = \texttt{waitFrames} + \texttt{Ntime} \times \texttt{tFactor}$, where \texttt{waitFrames} is how many time steps the program waits before it begins saving the field to the output array; \texttt{Ntime} is the user-input number of time-points, and \texttt{tFactor} is an integer with a default value of five. This is relevant to calculation of convergence conditions since the actual time step used in the simulation is therefore one fifth of the user input time step. Within each iteration of the loop, the four kernels corresponding to the four steps of the RK4 integration method are called sequentially. Each kernel is parallelized over the transverse $x$ and axial $z$ coordinates of the grid. 

Calculation of HHG propagation requires careful attention to convergence criteria. The RK4 solution used in this code is an explicit solver method and therefore is not unconditionally stable, which does mean that the program is more likely to diverge than produce physically nonsense results. To ensure convergence and minimize numerical dispersion, we require the distance traveled by the wave in one time step $dt$ to be less than or equal to the grid spacing along the propagation direction $dz$ so that the propagation obeys causality, i.e. our mesh must obey the Courant-Friedrichs-Lewy (CFL) condition. For a three-dimensional material model, this bounds the $dt$, $dz$, and $dx$ step sizes as:
\begin{equation}
v \Delta t \sqrt{\frac{1}{\Delta x^2}+\frac{1}{\Delta z^2}} \leq 1 
\end{equation}
In vacuum, $v=c$ the speed of light, but in a dispersive medium, $v = n(\omega) c$ and thus the convergence criterion naturally becomes frequency dependent. In the case of strong-field tunneling ionization, the material response produces a wave containing many frequencies which may cover a varying range of the dispersion curve for the refractive index. This introduces some numerical difficulty in ensuring sufficient step sizes for propagation of such a broadband pulse. 

\section*{Appendix B: Parameter space of microscopic simulations}
\setcounter{figure}{0}
\renewcommand{\thefigure}{B.\arabic{figure}}
\setcounter{equation}{0}
\renewcommand{\theequation}{B.\arabic{equation}}

The necessity of the dephasing time in producing distinct harmonic peaks in the plateau region of the spectra for the semiconductor material is consistent with prior literature \cite{Han2019ExtractionGeneration}.  We considered the effects of longitudinal dephasing $T_1$ of the carrier densities to damp asymmetric population oscillations but the found that the magnitude of $T_1$ has minimal effect on the output harmonic spectra, so we neglect it in our model.

\begin{figure}[H]
    \centering
    \includegraphics[width=1\linewidth]{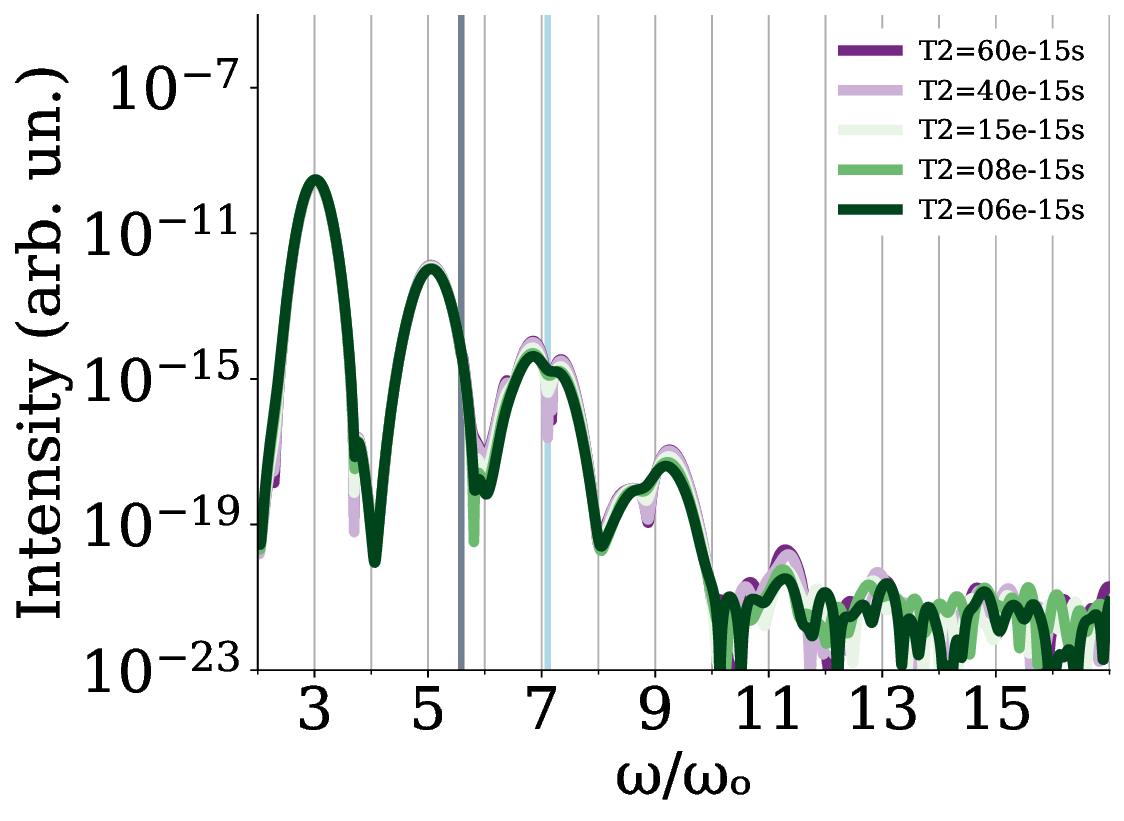} 
    \caption{$T_2$ analysis. The grey line refers to the direct band gap of the material, and the light blue line refers to maximum valence to conduction band energy gap described by our dispersion model. Driving field is centered at $\lambda_0=2.0 \mu m$, $E_0=6.63\times10^9$ $V/m$.} 
    \label{fig:placeholder}
\end{figure}

Figure B.1 shows the dependence of the spectra on the $T_2$ parameter from the 2.0 $\mu m$ driving laser of amplitude $6.62\times10^9$ $V/m$ using the SBEs in isolation (i.e. not coupled to any propagation equation), where the blue line corresponds to the maximum bandgap present in the two band dispersion curves described within the SBEs, and the dark grey line corresponds to the direct bandgap of $3.3$ eV. The spectra presented in these figures is calculated by: $S(\omega) = |P_{inter}(\omega) + i \omega J_{intra}(\omega)|^2$. It is clear that a cutoff appears around the maximum valence-to-conduction energy gap, and that shorter dephasing times produce slightly cleaner harmonic spectra. We note that for different input pulse conditions (in this case, $\lambda_0$=$3.5\mu m$ and a many-cycle pulse), the effect of $T_2$ is stronger and we find that shorter dephasing times of around 2.0-7.0 fs improve the noise floor of harmonics in the plateau region but also reduce the amplitude of below bandgap harmonics (Figure B.2)).

\begin{figure}[hbt!]
    \centering
    \includegraphics[width=1\linewidth]{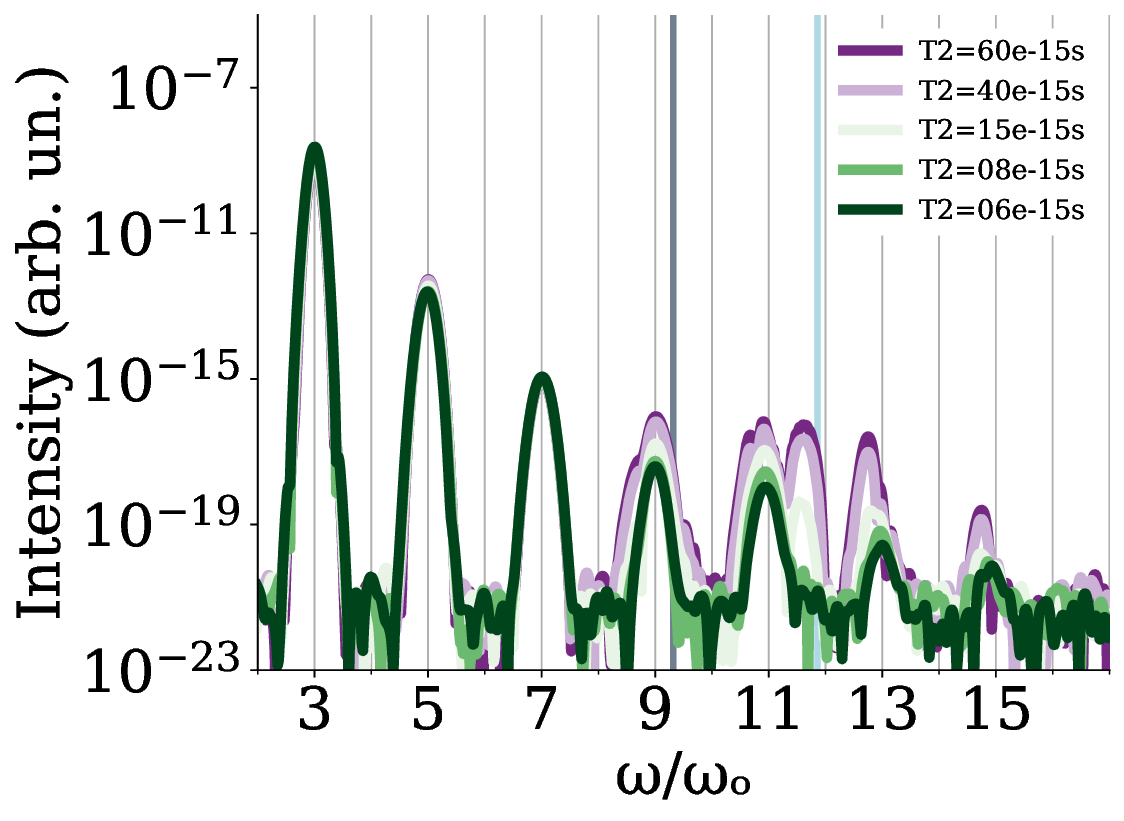} 
    \caption{$T_2$ scan of HHG spectra from a driving laser pulse with $\lambda_0$=$3.5\mu m$, $E_0=1.94\times10^9$ $V/m$, 83 fs duration.} 
    \label{fig:placeholder}
\end{figure}
We conclude from these microscopic simulations that $T_2 \leq 6$ fs is sufficient to produce numerically clean and replicable harmonic spectra for these input field parameters. Using a longer $T_2$ leads to increased high frequency components in the spectra, some of which can be attributed to numerical noise and thereby leads to unphysical high frequency components in the propagated pulse accumulating during propagation. Therefore, in order to reduce unphysical numerical noise in the simulated propagated spectra, we must use a sufficiently short dephasing time while calculating the SBE response in the context of this framework. 

Convergence tests for the mesh of timestep and $k$-kpoints shown in Figure B.3 were calculated in order to ensure that the timestep and $k$-points used in the propagated simulation were sufficient to resolve the microscopic SBE current. From these tests, it was concluded that any $dt < 100 as$ and $nk \geq 201$ would be sufficient to resolve all harmonics and minimize noise in the harmonic spectra. An odd number of k-points was used to ensure perfectly symmetric and periodic calculations over the Brillouin zone, and capture transitions at the $\Gamma$-point. 

\begin{figure}[hbt!]
    \centering
    \subfigure[]{
    \includegraphics[width=1\linewidth]{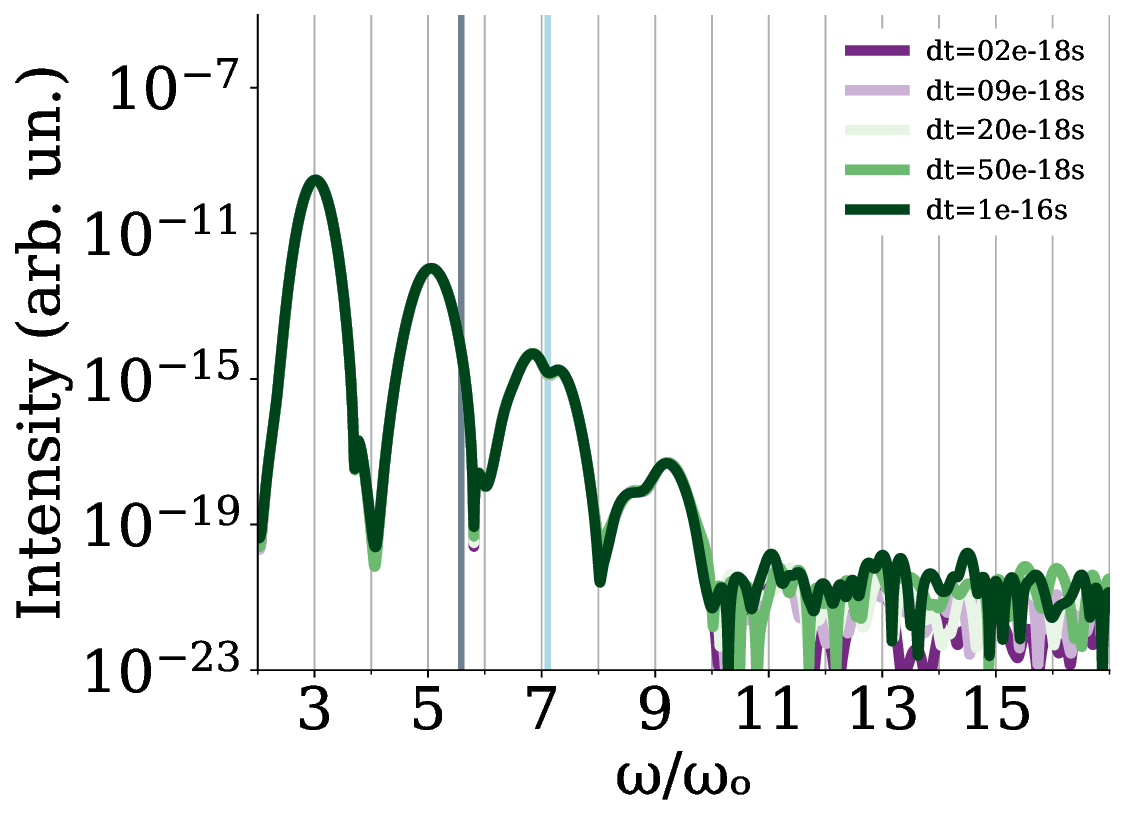}} 
    \subfigure[]{
    \includegraphics[width=1\linewidth]{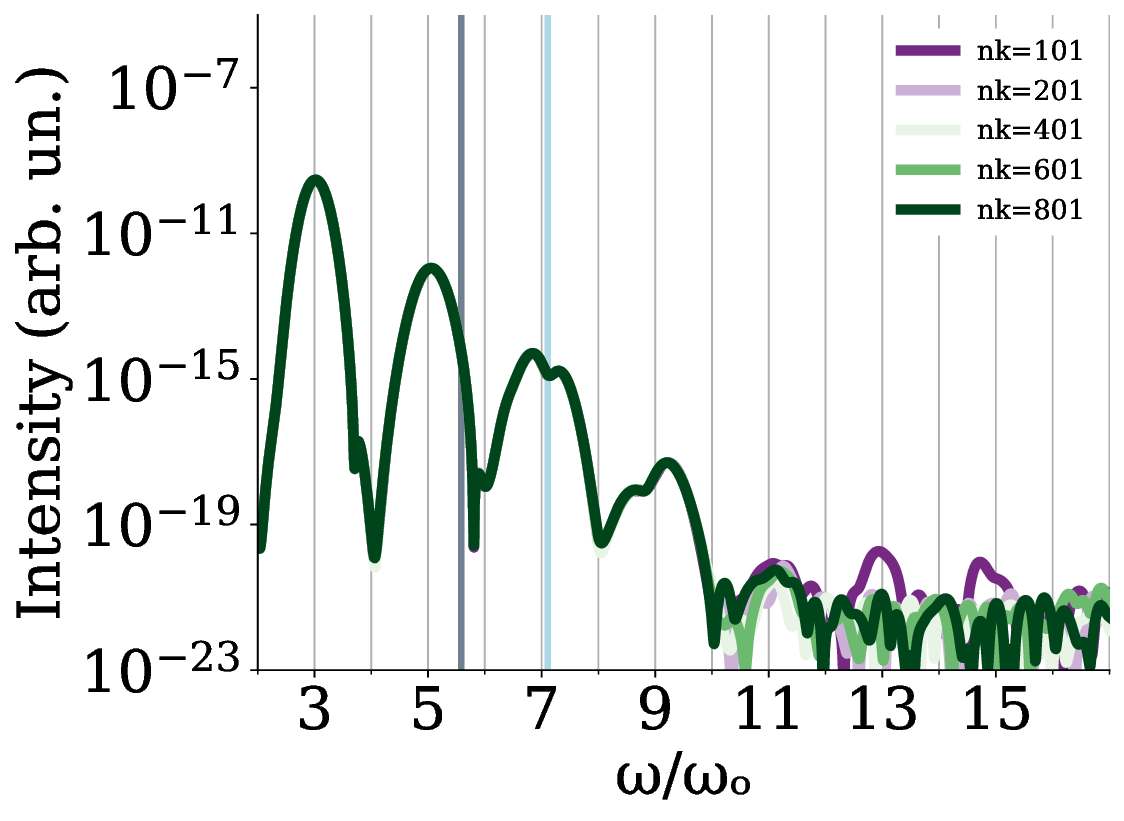}} 
    \caption{Convergence tests calculated from the total emitted current from SBEs (no propagation) varying (a) timestep, (b) number of $k$-points. Simulations use the same pulse duration, frequency, and amplitude as the propagation simulations shown elsewhere in this work. Default timestep was $9.0$ $as$, default $T_2=$ 8e-15 fs, and default $nk=200$.}
    \label{fig:placeholder}
\end{figure}

\section*{Appendix C: Tukey filtering in UPPE}
\setcounter{figure}{0}
\renewcommand{\thefigure}{C.\arabic{figure}}
\setcounter{equation}{0}
\renewcommand{\theequation}{C.\arabic{equation}}
We apply a Tukey filter ($\alpha=0.5$) to electric field waveform before each iteration SBE integration to reduce noise accumulated at the time domain edges from repeated FFTs, as shown in Figure C.1(a-c), to improve numerical convergence during propagation. As clear in Figure C.1(d), which is the normalized on-axis harmonic spectra, Tukey filtering reduces the noise floor but does leaves the on-axis spectral intensity of low frequency harmonics below the material bandgap intact, although above bandgap spectra may differ. In the UPPE code we therefore apply a Tukey filter before and after the computation of the SBE response within the nonlinear current calculation kernel during propagation.

\begin{figure*}[hbt!]
    \centering
    \subfigure[]{\includegraphics[width=0.45\linewidth]{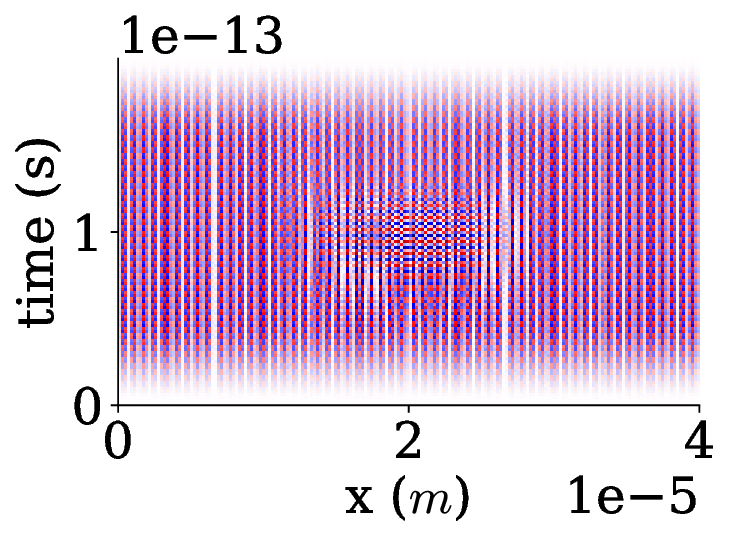}}
    \subfigure[]{\includegraphics[width=0.45\linewidth]{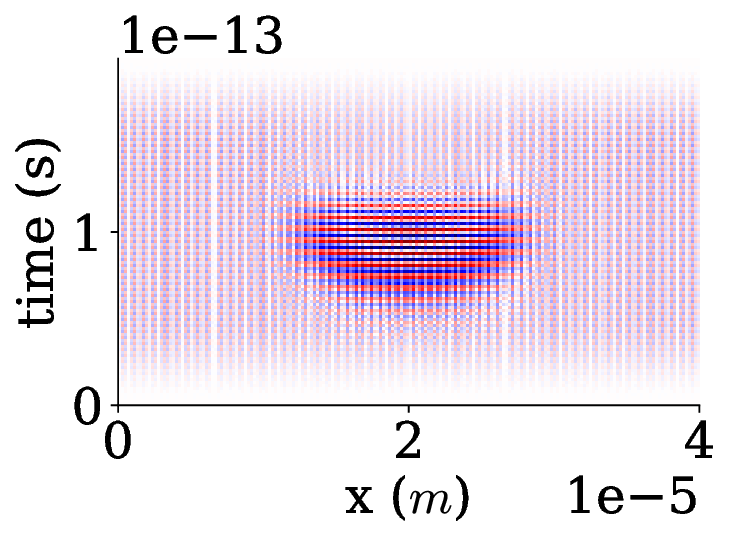}}
    \subfigure[]{\includegraphics[width=0.45\linewidth]{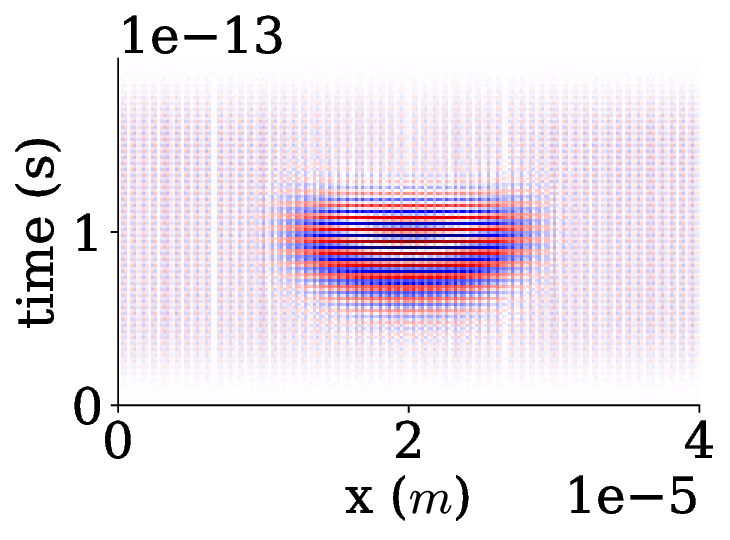}}
    \subfigure[]{\includegraphics[width=0.5\linewidth]{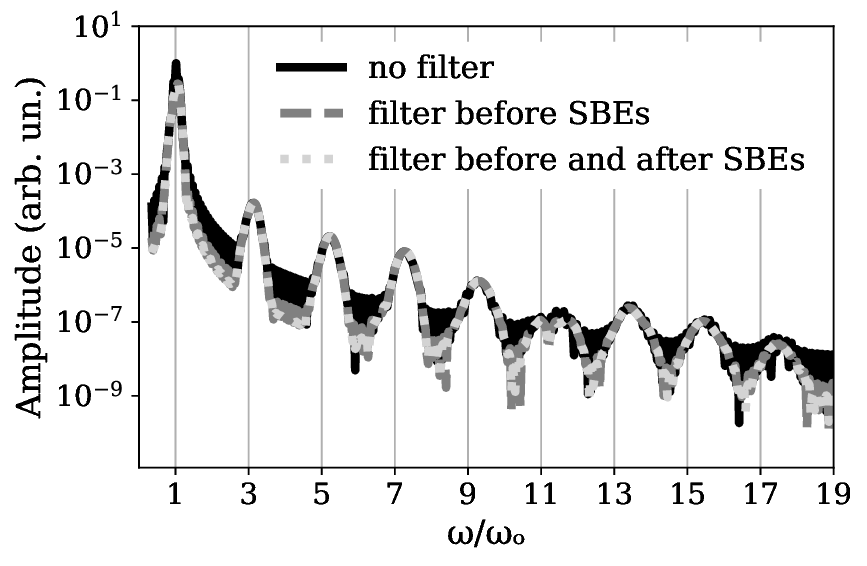}}
    \caption{Effect of applying Tukey filter within simulated propagation through 1 um thickness slab of a non-cenetrosymmetric semiconductor as evidenced by heatmaps of the transverse map of the pulse time signal (a) no Tukey filtering ("no filter"); (b) with Tukey filtering before each RK4 solution to the SBEs ("filter before SBEs") and (c) with Tukey filtering before and after each RK4 solution to the SBEs ("filter before and after SBEs"). (d) The on-axis harmonic trace resulting from each procedure. From these results, we apply the procedure depicted in (c) to produce cleanest UPPE propagated harmonic spectra.}
\end{figure*}

\section*{Appendix D: Semiconductor Bloch Equation Formulations}
\setcounter{figure}{0}
\renewcommand{\thefigure}{D.\arabic{figure}}
\setcounter{equation}{0}
\renewcommand{\theequation}{D.\arabic{equation}}
All equations are written in atomic units. Below we record the SBEs for an arbitrary number of bands expressed in the length gauge \cite{Luu2016High-orderApproach}. These include the equation of motion of the microscopic polarization between conduction and valence bands (Equation D.1), between conduction bands (Equation D.2), between valence bands (Equation D.3); and the equation of motion of the microscopic carrier densities of the electron (Equation D.4) and hole (Equation D.5):

\begin{strip}
\begin{align}
i \frac{\partial}{\partial t}p_k^{{h_i}{e_j}}  = & (\varepsilon_k ^{e_j} + \varepsilon_k^{h_i} - i/T_2) p_k^{{h_i}{e_j}} \nonumber - (1 - f^{e_j}_k - f^{h_i}_k)d_k^{{e_j}{h_i}}E(t) + iE(t)\nabla_k p_k^{{h_i}{e_j}}  \nonumber\\
& + E(t) \sum_{e_\lambda \neq e_j} (d_k^{{e_\lambda}{h_i}}p_k^{{e_\lambda}{e_j}} - d_k^{{e_j}{e_\lambda}}p_k^{{h_i}{e_\lambda}} + E(t)\sum_{h_\lambda \neq h_i}(d_k^{{h_\lambda}{h_i}}p_k^{{h_\lambda}{e_j}} - d_k^{{e_j}{h_\lambda}}p_k^{{h_i}{h_\lambda}})
\end{align}
\end{strip}
\clearpage

\begin{strip}
\begin{align}   
i \frac{\partial}{\partial t}p_k^{{e_i}{e_j}} = & (\varepsilon_k ^{e_j} -\varepsilon_k^{e_i} - i/T_2) p_k^{{e_i}{e_j}}\nonumber + (f^{e_j}_k - f^{e_i}_k)d_k^{{e_j}{e_i}}E(t) + iE(t)\nabla_k p_k^{{e_i}{e_j}} \nonumber \\ 
 & + E(t) \sum_{e_\lambda \neq e_j} d_k^{{e_\lambda}{e_i}} P_k^{{e_\lambda}{e_j}} - E(t) \sum_{e_\lambda \neq e_i} d_k^{{e_j}{e_\lambda}}p_k^{{e_i}{e_\lambda}} + E(t)\sum_{h_\lambda}(d_k^{{h_\lambda}{e_i}}p_k^{{h_\lambda}{e_j}} - d_k^{{e_j}{h_\lambda}})(p_k^{{h_\lambda}{e_i}})^*)  
\end{align}

\begin{align}
i \frac{\partial}{\partial t}p_k^{{h_i}{h_j}} =& (\varepsilon_k ^{h_i} - \varepsilon_k^{h_j} - i/T_2) p_k^{{h_i}{h_j}} + (f^{h_i}_k - f^{h_j}_k)d_k^{{h_j}{h_i}}E(t) + iE(t)\nabla_k p_k^{{h_i}{h_j}} + E(t) \sum_{h_\lambda \neq h_j} d_k^{{h_\lambda}{h_i}} P_k^{{h_\lambda}{h_j}} \nonumber\\
&- E(t) \sum_{h_\lambda \neq h_i} d_k^{{h_j}{h_\lambda}}p_k^{{h_i}{h_\lambda}}  + E(t)\sum_{e_\lambda}(d_k^{{e_\lambda}{h_i}}(p_k^{{h_j}{e_\lambda}})^* - d_k^{{h_j}{e_\lambda}}p_k^{{h_i}{e_\lambda}})
\end{align} 

\begin{align}
\frac{\partial}{\partial t} f_k^{e_i} = & \frac{\partial}{\partial t} f_k^{e_i} |_{relax} - 2Im[\sum_{e_\lambda \neq e_i} d_k^{{e_i}{e_\lambda}} E(t) (p_k^{{e_\lambda}{e_i}})^* + \sum_{h_\lambda} d_k^{{e_i}{h_\lambda}}E(t)(p_k^{{h_\lambda}{e_i}})^*] + E(t)\nabla_kf_k^{e_i} 
\end{align}

\begin{align}
\frac{\partial}{\partial t} f_k^{h_i} = &\frac{\partial}{\partial t} f_k^{h_i} |_{relax} - 2Im[\sum_{h_\lambda \neq h_i} d_k^{{h_\lambda}{h_i}} E(t) (p_k^{{h_i}{h_\lambda}})^* + \sum_{e_\lambda} d_k^{{e_\lambda}{h_i}}E(t)(p_k^{{h_i}{e_\lambda}})^*] + E(t)\nabla_kf_k^{h_i}
\end{align}
\end{strip}

\end{document}